\documentclass{llncs}
\usepackage{url}
\usepackage{citesort}
\usepackage{times}
\usepackage{amssymb}
\usepackage{mathpartir}
\usepackage{graphics}
\usepackage[all]{xypic}
\usepackage{enumerate}
\usepackage{color}
\usepackage{rotating}
\usepackage{esvect}
\usepackage{shading}

\newcommand{\orelse}{~|~}
\newcommand{\andalso}{~\wedge~}
\newcommand{\inferx}[2]{#1~\Rightarrow~ #2}
\newcommand{\setctabs}{\hspace*{2em}\=\hspace*{2em}\=\hspace*{2em}\=
  \hspace*{2em}\=\hspace*{8em}\= \kill}

\makeatletter 
\def\overbracket#1{\mathop{\vbox{\ialign{##\crcr\noalign{\kern3\p@} \downbracketfill\crcr\noalign{\kern3\p@\nointerlineskip} $\hfil\displaystyle{#1}\hfil$\crcr}}}\limits} 
\def\underbracket#1{\mathop{\vtop{\ialign{##\crcr $\hfil\displaystyle{#1}\hfil$\crcr\noalign{\kern3\p@\nointerlineskip} \upbracketfill\crcr\noalign{\kern3\p@}}}}\limits} \def\overparenthesis#1{\mathop{\vbox{\ialign{##\crcr\noalign{\ker
n
3\p@} \downparenthfill\crcr\noalign{\kern3\p@\nointerlineskip} $\hfil\displaystyle{#1}\hfil$\crcr}}}\limits} 
\def\underparenthesis#1{\mathop{\vtop{\ialign{##\crcr
        $\hfil\displaystyle{#1}\hfil$\crcr\noalign{\kern3\p@\nointerlineskip} \upparenthfill\crcr\noalign{\kern3\p@}}}}\limits} 
\def\downparenthfill{$\m@th\braceld\leaders\vrule\hfill\bracerd$} 
\def\upparenthfill{$\m@th\bracelu\leaders\vrule\hfill\braceru$} 
\def\upbracketfill{$\m@th\makesm@sh{\llap{\vrule\@height3\p@\@width.7\p@}}% 
\leaders\vrule\@height.7\p@\hfill
\makesm@sh{\rlap{\vrule\@height3\p@\@width.7\p@}}$} 
\def\downbracketfill{$\m@th \makesm@sh{\llap{\vrule\@height.7\p@\@depth2.3\p@\@width.7\p@}}% 
\leaders\vrule\@height.7\p@\hfill \makesm@sh{\rlap{\vrule\@height.7\p@\@depth2.3\p@\@width.7\p@}}$}
\def\equalsfill{$\m@th\mathord=\mkern-7mu \cleaders\hbox{$\!\mathord=\!$}\hfill \mkern-7mu\mathord=$}  
\makeatother 

\newcommand{\cenv}[3]{
\begin{flushleft}
\parbox{12.2cm}{{\bf #1} $~#2$}
\\
\vspace{-1mm}
\parbox{12.2cm}{\downbracketfill}
\\
\vspace{-3mm}
\end{flushleft}
#3
\begin{flushleft}
\vspace{-2mm}
\parbox{12.2cm}{\upbracketfill}
\end{flushleft}
}

\newcommand{\compactcenv}[1]{
\begin{flushleft}
\parbox{12.2cm}{\downbracketfill}
\\
\vspace{-3mm}
\end{flushleft}
#1
\begin{flushleft}
\vspace{-3mm}
\parbox{12.2cm}{\upbracketfill}
\end{flushleft}
}

\newcommand{\cenvv}[3]{\cenv{#1}{#2}{#3}}

\newcommand{\abs}[2]{\lambda #1.~#2}
\newcommand{\app}[2]{(#1~#2)}
\newcommand{\bind}[3]{\mathsf{bind}~#1 = #2~\mathsf{in}~#3}
\newcommand{\strong}[2]{(\eta_{#1}~#2)}
\newcommand{\weak}[2]{(\overline \eta_{#1}~#2)}

\newcommand{\tup}[2]{\langle #1, #2 \rangle}
\newcommand{\proj}[2]{(\mathsf{proj}_{#1}~#2)}
\newcommand{\inj}[2]{(\mathsf{inj}_{#1}~#2)}
\newcommand{\split}[4]{\mathsf{case}~#1~\mathsf{of}~\mathsf{inj}_1(#2).~#3~\|~\mathsf{inj}_2(#2).~#4}
\newcommand{\weaken}[1]{(\mathsf{weaken}~#1)}
\newcommand{\weakbind}[3]{\mathsf{weakbind}~#1 = #2~\mathsf{in}~#3}

\newcommand{\unitT}{\mathsf{unit}}
\newcommand{\unit}{\mathsf{()}}

\newcommand{\dcc}{DCC}
\newcommand{\dccd}{DCC${}^{\it d}$}
\newcommand{\dccdc}{DCC${}^{\it dc}$}
\newcommand{\dcccd}{DCC${}^{\it cd}$}

\newcommand{\trule}[1]{$({\sf T\mbox{-}#1})$}

\title{Liberalizing Dependency}
\author{Avik Chaudhuri}
\institute{University of Maryland at College Park \\
\url{avik@cs.umd.edu}}

\begin{document}
\maketitle
\begin{abstract}
%The dependency core calculus (\dcc) is an extension of the computational lambda calculus, designed to capture a central notion of dependency common to many programming language settings, including security. This notion of dependency tracks not only explicit effects due to data flow, but also implicit effects due to  control flow. In this paper, we deconstruct \dcc~and study variants in which explicit and implicit effects are decoupled. This allows us to consider settings where a weaker notion of dependency---one that tracks only explicit effects due to data flow---may coexist with \dcc's stronger notion of dependency, safely and symbiotically. In particular, we show for the first time how strong, noninterference-based security can be reconciled with weak, trace-based security within the same system. 
The dependency core calculus (\dcc), a simple extension of the computational lambda calculus, captures a common notion of dependency that arises in many programming language settings. This notion of dependency is closely related to the notion of information flow in security; it is sensitive not only to data dependencies that cause explicit flows, but also to control dependencies that cause implicit flows. In this paper, we study variants of \dcc~in which the data and control dependencies are decoupled. This allows us to consider settings where a weaker notion of dependency---one that restricts only explicit flows---may usefully coexist with \dcc's stronger notion of dependency. In particular, we show how strong, noninterference-based security may be reconciled with weak, trace-based security within the same system, enhancing soundness of the latter and completeness of the former. 
\end{abstract}

\pagestyle{plain}
\thispagestyle{plain}

\section{Introduction}\label{sec:intro}
The dependency core calculus (\dcc) \cite{abadi1999core} is a simple extension of the computational lambda calculus \cite{moggi1991notions}, where each level $\ell$ in a lattice is associated with a type constructor $T_\ell$ that behaves as a monad. \dcc~was designed to capture a central notion of dependency common to many programming language settings, including security. This notion of dependency is closely related to the concepts of parametricity \cite{reynolds1983types,tse2004translating} and noninterference \cite{goguen1982security,abadi1999core}. Roughly, \dcc's type system guarantees that the computational effects of a program protected by some level $\ell$ can only be observed by programs protected by levels $\ell$ or higher in the lattice. Of course, such effects may include not only explicit effects due to data flow, but also implicit effects due to  control flow. For example, consider the following functions.
\begin{eqnarray*}
f & = & \abs {x: T_\ell(s_1 + s_2)} \bind y x y \\
g & = & \abs {x: T_\ell(s_1 + s_2)} \bind y x \split y z {\inj 1 \unit} {\inj 2 \unit}
\end{eqnarray*}
The type of the argument $x$ is an $\ell$-protected sum type $(s_1 + s_2)$, denoted $T_\ell(s_1 + s_2)$. A value of this type is of the form $\strong \ell {\inj i {e_i}}$, $i \in \{1,2\}$, where $e_i$ is an expression of type $s_i$, $\mathsf{inj}_i$ is a case constructor, and $\eta_\ell$ denotes some $\ell$-protection mechanism (which can be undone with $\mathsf{bind}$). The function $f$ undoes the protection on $x$ and returns it. The function $g$ also undoes the protection on $x$, but returns only its case constructor. Of course, neither function is typable in \dcc, since $f$ and $g$ return unprotected results that depend on $x$---in other words, $f$ and $g$ leak information on $x$.
%
%\begin{eqnarray*}
%f & : & T_\mathtt{high}(s + s) \rightarrow (s + s) \\
%g & : & T_\mathtt{high}(s + s) \rightarrow (\unit + \unit)
%\end{eqnarray*}
%
Still, intuitively $g$ may seem ``safer'' than $f$---while $f$ explicitly reveals all information on $x$ through data flow, $g$ implicitly reveals only one bit of information on $x$ through control flow. %(Of course, this advantage vanishes if $x$ contains only one bit of information, such as if $x$ is a boolean; on the other hand, if $x$ is a list of integers, the advantage may be significant---while $f$ reveals the list itself, $g$ only reveals whether the list is empty.)

Traditionally, security experts have dismissed this notion of ``safety'' as unsound, since the attacker might be able to amplify the one-bit leak of information in $g$ to leak all information on $x$, thereby making it as dangerous as $f$. However,  for non-malicious code, such attacks are often complex and seem rare in practice \cite{king2008implicit}. Indeed, in the past few years several static analyses for security have focused on restricting effects due to data flow, while ignoring other effects~%abadi02sectyplog,cardelli05secgp,
\cite{clause07dytan,shankar06automated,suh04secure,yin07panorama,costa05vigilante,castro06dataflow,vogt07xss,dalton07raksha,broadwell03scrash,chen07format,martin05sql,shankar2001detecting,xie07saturn,zhang02cqual,chaudhuri2009type,tripp2009taj,chaudhuri2009language}. From a theoretical perspective, one may simply consider these analyses unsound, and assume that they provide no guarantee. Alternatively, one may try to understand the precise guarantee that these analyses provide, and evaluate whether such a guarantee is at all important for security. This is the stance we take in this paper.

Previous work on downgrading and robustness \cite{zdancewic2001robust,myers2004enforcing} is based on similar concerns. Roughly, downgrading allows some specific information in the system to be released, and robustness guarantees that this does not cause further, unintentional leak of information in the system. For example, a function $p$ that checks whether a given password is correct releases information on the correct password whenever it returns the result of the check. A system using $p$ may still be robust, in the sense that the attacker cannot exploit the information released by $p$ to leak further information in the system.

However, downgrading as a mechanism of information release may be too coarse. For example, it blurs the qualitative distinction between a usual password-checking function that releases partial information on the correct password, and a function that releases the correct password itself. (This distinction is similar to the one between functions $g$ and $f$ above.) In this paper, we explore a finer mechanism of information release, called \emph{weakening}. In particular, weakening the protection on the correct password allows information on it to be released implicitly through control flow, but not explicitly through data flow. As usual, robustness may still require that such weakening does not trigger further weakening in the system. %In this setting, \emph{robustness} requires that weakening at level $\ell$ is not influenced by weakening at levels lower than $\ell$.

We study weakening and its properties by considering variants of \dcc~in which explicit and implicit effects are decoupled. The implicit effects arise entirely out of case analysis, so the main differences with \dcc~lie in the handling of sum types. For instance, consider the following typing rule in \dcc:
$$\infer
	{\Gamma \vdash e : T_\ell (s) \\ \Gamma, x : s \vdash e' : t \\ \mbox{$t$ is protected at $\ell$}}
	{\Gamma \vdash (\mathsf{bind}~x = e~\mathsf{in}~e') : t}
$$
The variable $x$ binds the result of $e$ upon undoing its protection. Since $x$ is in the scope of $e'$, the computational effects of $e'$ should only be observable to programs that are protected by levels $\ell$ or higher. This is ensured by the side condition, which restricts $t$ to be only of certain forms. (We will review the formal definition of this condition later.) In particular, $t$ cannot be a sum type, because information on $x$ may be leaked through the case constructor of a value of such type.  Indeed, this is exactly why $f$ and $g$ are not typable in \dcc; their results have, respectively, types $(s_1 + s_2)$ and $(\unitT + \unitT)$.

In contrast, we study the following typing rule in  \dccd, a variant of \dcc:
$$\infer
	{\Gamma \vdash e : \overline T_\ell (s) ~\quad \Gamma, x : s^\ell \vdash e' : t \\ \mbox{$t$ is weakly protected at $\ell$}}
	{\Gamma \vdash (\mathsf{bind}~x = e~\mathsf{in}~e') : t}
$$
The type constructor $\overline T_\ell$ provides weaker protection than $T_\ell$; it focuses on restricting effects due to data flow, while ignoring other effects. %A value of this type has the form $\overline\eta_\ell(e'')$, where $e''$ is of type $s$. 
In particular, the side condition in the rule above allows $t$ to be a sum type. At the same time, $t$ is adequately restricted to ensure that $x$ itself is not released without protection. We introduce \emph{open types} for this purpose; roughly, an open type $s^\ell$ is given to a value of type $s$ that requires weak protection by level $\ell$. We assume such a type for $x$, and prevent $t$ from being an open type. In the resulting system, $g$ is typable (after weakening the type of the argument) but $f$ is not. We show that if a program is typable in \dcc, then it remains typable in \dccd~by weakening types. 
Furthermore, we formalize the precise guarantee enforced by \dccd. This guarantee is related to Volpano's definition of \emph{weak security} as a safety property~\cite{volpano99sas}, and it eliminates (at least) Denning and Denning's \emph{explicit flow} attacks~\cite{denning1977certification}. For non-malicious code, \emph{i.e.}, code that the attacker cannot fully control, such attacks are far more dangerous than implicit flow attacks \cite{russo09implicit}; thus \dccd's guarantee is important for security of such code, at least from a practical perspective. 

While the typing rules of \dccd~have an interesting flavor of their own, mixing them with \dcc's typing rules can yield surprisingly pleasant cocktails. We explore a couple of such recipes in this paper; they highlight the symbiotic nature of these systems. %, that are not only safe but also symbiotic.  
\begin{itemize}
\item
We study a dynamic $\mathsf{weaken}$ primitive that allows values of type $T_\ell$ to be cast as values of type $\overline T_\ell$. This weakening may invalidate strong protection guarantees at levels $\ell$ and lower. However, weak protection guarantees should still hold at these levels, and strong protection guarantees should hold at all other levels. We show how these guarantees can be enforced by recycling \dcc's types to carry \emph{blames} for weakening. Specifically, we include the rule
$$\infer
	{\Gamma \vdash e : T_\ell (s)}
	{\Gamma \vdash \mathsf{weaken}~e : T_{\beta(\ell)} (\overline T_\ell (s))}
$$
where $\beta$ is some isomorphism from the lattice of levels to some lattice of blames. 
The behavior of the resulting system, \dccdc, rests on the definition of $\beta$. 
\begin{itemize}
\item If $\beta$ preserves joins and meets, then a program's type carries a blame $\beta(\ell)$ such that $\ell$ \emph{upper-bounds} the levels of weakening on which its results may depend; thus, strong protection guarantees continue to hold at all levels not $\ell$ or lower. 
\item If $\beta$ exchanges joins and meets, then a program's type carries a blame $\beta(\ell)$ such that $\ell$ \emph{lower-bounds} the levels of weakening on which its results may depend; thus, weak protection guarantees are robust against all levels not $\ell$ or higher.
\end{itemize}
\item 
Conversely, we show how a \dccd-style analysis can make \dcc's dependency analysis more precise. Consider the following functions, which are clearly secure yet rejected by DCC because sum types are never considered protected: 
$$\abs x {\bind y x {\inj i \unit}}\qquad i \in \{1,2\}$$
%
%Indeed, the restriction on sum types is too harsh. 
To typecheck such functions, we observe that any information leak is ultimately due to either an explicit leak through data flow or an implicit leak through control flow. Specifically, evaluating an expression of sum type may reveal information about sensitive data only if that expression either does a case analysis on sensitive data, or releases the sensitive data itself. We can prevent the former possibility by including a side condition in the rule for $\mathsf{case}$, and the latter by delegating to \dccd's typing rules. We show that the resulting system, \dcccd, is sound and more liberal than \dcc; in particular, it admits some new type-preserving optimizations.
\end{itemize}

\noindent
In the context of security, these results suggest some interesting ways in which strong, noninterference-based security may be reconciled with weak, trace-based security within the same system, enhancing soundness of the latter and completeness of the former. Specifically, in a system where protection may have been partially weakened, a strong blame analysis can be used to provide strong protection guarantees for those parts of the system that are not affected by such weakening. Conversely, a weak flow analysis can be used to increase the coverage of such guarantees. 

To summarize, we make the following contributions in this paper.
\begin{itemize}
\item We deconstruct \dcc, which captures standard information flow, into a weaker system \dccd~that is instead focused on explicit information flow. We argue that this system provides the foundations for several recent static analyses for security that do not restrict implicit information flow (Section \ref{sec:dccd}).
\item We study a language primitive $\mathsf{weaken}$ that switches from \dcc-style protection to \dccd-style protection of programs at run time. Such weakening may be viewed as a milder form of downgrading that preserves data-flow guarantees for the resulting programs. Furthermore, we show how such weakening can be controlled by reusing \dcc~mechanisms to associate blames for weakening (Section \ref{sec:dccdc}).  
\item Going in the other direction, we study how \dccd's typing rules can enhance the precision of \dcc's typing rules. This technique (once again) relies on deconstructing information flow into explicit and implicit information flow (Section \ref{sec:dcccd}).
\end{itemize}
%
%Overall, we believe that the main importance of these results lies in their conceptual rather than technical details. Still, the technical details may be of interest. For instance, we need to develop a theory of open types to correctly track explicit flows in \dccd~(Section \ref{sec:dccd}). Mixing the typing rules of \dcc~and \dccd~to obtain improved hybrid systems (Sections \ref{sec:dccdc} and \ref{sec:dcccd}) also requires much care for correctness.

We review \dcc~next (Section \ref{sec:dcc}), deferring further discussion on  related work and conclusions until the end (Section \ref{sec:relwork}). 

%Talk about implicit trust in case analysis. 

\section{Background on \dcc}\label{sec:dcc}

Recall that the computational lambda calculus \cite{moggi1991notions} extends the simply typed lambda calculus with a type constructor that is interpreted as a monad. The monad is used to systematically control effects in the language. %; the same idea appears in Haskell to guarantee that ``pure'' functions cannot depend on ``impure'' computations. 
\dcc~\cite{abadi1999core} carries this idea further by distinguishing computations at various ``levels'', and controlling effects across levels. Specifically, \dcc~includes a monadic type constructor for each level in a lattice, and has a special typing rule that restricts how computations at various levels may be composed based on the lattice. %We review this system below.
Let $\ell$ denote levels in such a lattice with ordering $\sqsubseteq$, join $\sqcup$, meet $\sqcap$, bottom $\bot$, and top $\top$. We focus on the following syntax for types and terms in \dcc. (For brevity, we omit any discussion of pointed types and recursive programs; see Section \ref{sec:relwork} for further comments.)  \\
\vspace{-7mm}
\compactcenv{
\vspace{-1mm}
%\line(1,0){348}\\
\emph{types} $s, t ::= \unitT \orelse (s \rightarrow t) \orelse (s \times t) \orelse (s + t) \orelse T_\ell(s)$ \\ 
%\vspace{-3mm} \\
\emph{values} $v ::= \unit \orelse \abs x e \orelse \tup e {e'} \orelse \inj i e \orelse \strong \ell e$\\
%\vspace{-3mm} \\
\emph{terms} $e ::= v \orelse \app e {e'} \orelse \proj i e \orelse (\split e x {e_1} {e_2}) \orelse (\bind x e {e'})$
%\vspace{-3mm} \\
%\emph{terms} $e,v ::= \unit \orelse \abs x e \orelse \app e {e'} \orelse \tup e {e'} \orelse \proj i e$\\
%$~~\quad\qquad\qquad  \orelse \inj i e \orelse (\split e x {e_1} {e_2}) \orelse \strong \ell e \orelse (\bind x e {e'})$
%\vspace{-3mm} \\
%\emph{terms} $e,v ::= \unit \orelse \abs x e \orelse \app e {e'} \orelse \tup e {e'} \orelse \proj i e$\\
%$~~\quad\qquad\qquad  \orelse \inj i e \orelse (\split e x {e_1} {e_2}) \orelse \strong \ell e \orelse (\bind x e {e'})$
}
%\vspace{-1mm} \\
%\line(1,0){348}
%

Types include unit, product, sum, and function types, as well as types $T_\ell(s)$ for each level $\ell$ in the lattice. Terms include the introduction and elimination forms for these types; the introduction forms are considered values.  In particular, $\strong \ell e$ has type $T_\ell(s)$ whenever $e$ has type $s$, and $(\bind x {\strong \ell e} {e'})$ reduces to $e'[e/x]$.

In practice, $\eta_\ell$ may represent any mechanism that provides ``protection'' at level $\ell$, broadly construed. In the context of secrecy, for instance, $\strong \ell e$ may be viewed as an encryption of $e$ with a key secret to level $\ell$. %(Conversely, in the context of integrity, $\strong \ell e$ may be viewed as a copy of $e$ that is tainted by $\ell$.) 
The typing rule for $\mathsf{bind}$ should then ensure that the secrecy of $e$ is preserved in the above reduction. In particular, this may require that the result be similarly encrypted. 
This intuition is captured by a predicate $\ell \preceq t$, read as ``$t$ is protected at $\ell$'', meaning that terms of type $t$ cannot leak any information at level $\ell$---in other words, terms of type $t$ are indistinguishable to any level $\ell'$ that is not at least $\ell$ in the lattice. The following rules define this predicate:
$\ell \preceq \unitT$; $\ell \preceq (s \rightarrow t)$ iff $\ell \preceq t$; $\ell \preceq (s \times t)$ iff $\ell \preceq s$ and $\ell \preceq t$; and $\ell \preceq T_{\ell'}(s)$ iff $\ell \sqsubseteq \ell'$ or $\ell \preceq s$. 
%
%\cenv{Protection rules}{}{
%\vspace{-3mm}
%\[
%\left.
%\begin{array}{rl}
%{\sf (P\mbox{-}unit)}~ & {\ell \preceq \unitT}
%\\ 
%{\sf (P\mbox{-}function)}~ & \inferx
%	{\ell \preceq t}
%	{\ell \preceq (s \rightarrow t)}
%\\
%{\sf (P\mbox{-}product)}~ & \inferx
%	{\ell \preceq s \andalso \ell \preceq t}
%	{\ell \preceq (s \times t)}
%\\ 
%{\sf (P\mbox{-}monad\mbox{-}1)}~ & \inferx
%	{\ell \sqsubseteq \ell'}
%	{\ell \preceq T_{\ell'}(s)}
%\\	
%{\sf (P\mbox{-}monad\mbox{-}2)}~ & \inferx
%	{\ell \preceq s}
%	{\ell \preceq T_{\ell'}(s)}
%\end{array}
%\right.
%\]
%}
%\vspace{-1mm}
%
%These rules may be explained as follows. The term $\unit$ of type $\unitT$ cannot leak any information since it is the only term of that type. Terms of type $(s \times t)$---which evaluate to tuples---cannot leak any more information than their projections, which have types $s$ and $t$. Similarly, terms of type $(s \rightarrow t)$---which evaluate to functions---cannot leak any more information than their bodies, which have type $t$. Finally, terms of type $T_{\ell'}(s)$---which evaluate to terms protected at level $\ell'$---cannot leak information at level $\ell$ if $\ell'$ is at least $\ell$; and in any case, they cannot leak any more information than their unprotected payloads.

Significantly, this definition does not consider sum types to be protected. The broad reason is that any information in terms is ultimately conveyed by case constructors. (The other constructors---unit, tupling, function abstraction, and $\ell$-protection---cannot convey any information since they are completely determined by the associated types.) For instance, a boolean may be encoded as either $\inj 1 \unit$ or $\inj 2 \unit$, thereby conveying one bit of information; so the sum type $(\unitT + \unitT)$ can serve as an encoding of the datatype $\mathtt{boolean}$. %Similarly, $(\unitT + (\unitT + \unitT))$ can serve as an encoding of $\mathsf{boolean}~\mathsf{option}$, with $\inj 1 \unit$ encoding $\mathsf{none}$ and $\inj 2 b$ encoding $\mathsf{some}(b)$ for a boolean $b$. 
In general, complex datatypes can be encoded using sum types, and the only way of distinguishing terms of such types is by analyzing the case constructors used in those terms. Thus, it makes sense to require explicit protection on any term of a sum type. (However, we will show in Section \ref{sec:dcccd} that this restriction can be relaxed.) %(However, note that we will challenge this point in Section 5.)

%So what does $\ell \preceq t$ really mean? Literally, it means that $t$'s protection is at least $\ell$. This means that any information in $t$ must be protected by at least $\ell$, so that terms of type $t$ are indistinguishable to any level $\ell'$ that is not at least $\ell$ (\emph{i.e.}, $\ell \not\sqsubseteq \ell'$).

%Conversely, $\ell \not\preceq t$ means that it may be possible for some level $\ell'$ that is lower than or incomparable to $\ell$ to distinguish terms of type $t$. 

%\begin{theorem}
%If $\delta \vdash \ell \preceq s$ and $\ell' \sqsubseteq \ell$ then $\delta \vdash \ell' \preceq s$.
%\end{theorem}

The typing rules for \dcc~derive judgments of the form $\Gamma; \Pi \vdash e : t$, where $\Gamma$ contains type hypotheses for free variables and $\Pi$ is a \emph{protection context} \cite{tse2004translating}, %\footnote{Protection contexts did not appear in the original definition of \dcc~\cite{abadi1999core}, but their inclusion has some pleasant consequences; see \cite{tse2004translating}.}  
which indicates the maximum level of protection promised by the context. If $e$ is closed, $\Gamma$ is empty and $\Pi$ is $\bot$, and we use the simpler notation $\vdash e : t$ for the typing judgment. 
In addition to standard rules for the simply typed lambda calculus with sum and product types (see the appendix), we have: 
\vspace{-2mm}
\[ 
\left.
\begin{array}{rc}
{\sf (T\mbox{-}ret)} &\quad \infer
	{\Gamma; \Pi \sqcup \ell \vdash e : s}
	{\Gamma; \Pi \vdash \strong \ell e : T_\ell (s)}
\\  \vspace{-3mm}
\\
{\sf (T\mbox{-}bind)} &\quad \infer%*[sep=2mm]
	{\Gamma; \Pi \vdash e : T_\ell (s) \\ \Gamma, x : s; \Pi \vdash e' : t \\ \ell \preceq T_\Pi(t)}
	{\Gamma; \Pi \vdash \bind x {e} {e'} : t}
\end{array}
\right.
\]
\trule{ret} states that $\strong \ell e$ has type $T_\ell(s)$ whenever $e$ has type $s$, assuming $\ell$-protection by the context (as promised by joining $\ell$ with the protection context). \trule{bind} states that $(\bind x e {e'})$ has type $t$ whenever $e$ has a type of the form $T_\ell(s)$, $e'$ has type $t$ assuming that $x$ has type $s$, and the type $T_\Pi(t)$ is protected at $\ell$. The latter condition means that either $t$ is protected at $\ell$, or $\Pi$ is at least $\ell$; this ensures that the result of $e'$ cannot leak any information at $\ell$, including any information on $x$, which is bound to the result of $e$ upon undoing its $\ell$-protection at run time.

%\textshade[.91]{sharpcorners}{\gdef\outlineboxwidth{0.01}#1}

%\cenv{Figure \ref{fig:dccreln}: Indistinguishability relation (\dcc)}{}{
%\[
%\left.
%\begin{array}{lc}
%{\sf (I\mbox{-}unit)} & \infer
%	{}
%	{\bullet \sim_\ell \bullet : 1}
%\\ 
%{\sf (I\mbox{-}pair)} & \infer
%	{e_1 \sim_\ell e'_1 : s_1 \\ e_2 \sim_\ell e'_2 : s_2}
%	{\langle e_1, e_2\rangle \sim_\ell \langle e'_1, e'_2\rangle : s_1 \times s_2}
%\\ 
%{\sf (I\mbox{-}inj)} & \infer
%	{e_i \sim_\ell e'_i : s_i}
%	{\iota_i e_i \sim_\ell \iota_i e'_i : s_1 + s_2}
%\\ 
%{\sf (I\mbox{-}abs)} & \infer
%	{\forall e, e'.~~e \sim_\ell e' : s ~\Rightarrow~ (v~e) \sim_\ell (v'~e') : t}
%	{v \sim_\ell v' : s \rightarrow t}
%\\ 
%{\sf (I\mbox{-}ret)} & \infer
%	{\ell' \not\sqsubseteq \ell \\ \ell' \mbox{ not supplementary}}
%	{v \sim_\ell v' : \Sigma_{\ell'}(s)}
%\qquad \infer
%	{e \sim_\ell e' : s}
%	{(\eta_{\ell'}~e) \sim_\ell (\eta_{\ell'}~e') : \Sigma_{\ell'}(s)}
%\\ 
%{\sf (I\mbox{-}res)} & \infer
%	{e \sim_\ell e' : s}
%	{(\nu_{\ell'}~e) \sim_\ell (\nu_{\ell'}~e') : \mathcal B_{\ell'}(s)}
%\\ 
%{\sf (I\mbox{-}at)} & \infer
%	{v \sim_\ell v' : s}
%	{v \sim_\ell v' : s @ \ell}
%\\ 
%{\sf (I\mbox{-}eval)} & \infer
%	{e \longrightarrow^\star v \\ e' \longrightarrow^\star v' \\ v \sim_\ell v' : s}
%	{e \sim_\ell e' : s}
%\end{array}
%\right.
%\]
%}

The key property of this type system---ensuring a form of parametricity \cite{reynolds1983types,tse2004translating} or noninterference \cite{goguen1982security,abadi1999core}---can be formalized using a type-directed indistinguishability relation over terms, $e \sim_\ell e' : s$, meaning that terms $e$ and $e'$ of type $s$ are indistinguishable to level $\ell$. In addition to standard rules for logical equivalence (see the appendix), we have that $\strong {\ell'} e$ and $\strong {\ell'} {e'}$ are indistinguishable to $\ell$ unless $\ell$ is at least $\ell'$. 
In other words, we have that the encryptions of $e$ and $e'$ with a secret key at level $\ell'$ are indistinguishable to an observer at level $\ell$ as long as $\ell$ does not know any secrets at $\ell'$. 

For example, let $e_1 = \strong \ell {\inj 1 \unit}$ and $e_2 = \strong \ell {\inj 2 \unit}$, and suppose that $\ell \not\sqsubseteq \ell'$. Then $e_1 \sim_{\ell'} e_2 : T_\ell(\unitT + \unitT)$. Now recall the functions $f$ and $g$ defined in Section \ref{sec:intro} (and assume that $s_1 = s_2 = \unitT$ for simplicity). Then $f \not\sim_{\ell'} f$, since $\app f {e_1}$ reduces to $\inj 1 \unit$, $\app f {e_2}$ reduces to $\inj 2 \unit$, and $\inj 1 \unit \not\sim_{\ell'} \inj 2 \unit$. Similarly, we can show that $g \not\sim_{\ell'} g$. Fortunately, neither function is typable in \dcc. 
Next consider $f' = \abs x {\strong \ell {\app f x}}$ and $g' = \abs x {\strong \ell {\app g x}}$. Then we can show that $f' \sim_{\ell'} f' : T_\ell(\unitT + \unitT) \rightarrow T_\ell(\unitT + \unitT)$ and $g' \sim_{\ell'} g' : T_\ell(\unitT + \unitT) \rightarrow T_\ell(\unitT + \unitT)$, and both functions are typable in \dcc.
Indeed, the type system guarantees that whenever a typed function is applied to $\ell$-protected inputs, it always produces outputs that are indistinguishable to levels that are not at least $\ell$. 

\begin{theorem}[\dcc~soundness, \emph{cf.} \cite{tse2004translating}]\label{thm:dcc}
If $~\vdash e : T_\ell(s) \rightarrow t$, $~\vdash e_1 : s$, and $~\vdash e_2 : s$, then for any $\ell'$ such that $\ell \not\sqsubseteq \ell'$, $\app e {\strong \ell {e_1}} \sim_{\ell'} \app e {\strong \ell {e_2}} : t$. 
\end{theorem}

%
%This guarantee is a form of noninterference \cite{}, and is also related to the notion of parametricity \cite{}.

%\pagebreak

%The lattice of supplementary levels supplements $\sqsubseteq$ and $\sqcup$. 
%$$\infer
%	{\ell \sqsubseteq \ell'}
%	{\overline{\ell'} \sqsubseteq \overline \ell}
%\qquad \infer
%	{}
%	{\overline{\ell'} \sqcup \overline{\ell} = \overline{\ell \sqcap \ell'}}
%$$

%Roughly, the ``blame" of downgrading is tracked in a monad; the blame persists if it influences downgrading of higher levels, and it is transferred if it influences downgrading of lower levels. (This doesn't sound quite right; transferring the blame in any direction seems dangerous, unless we interpret it as ``trust".)

%Robustness: passive attacker as powerful as active attacker. 

\section{Explicit flows and \dccd}\label{sec:dccd}

While \dcc~can adequately encode various analyses, %in practice it is often difficult to satisfy, as argued in Section \ref{sec:intro}. Indeed, 
the underlying notion of dependency can be overly sensitive in certain settings. In this section, we design a variant of \dcc~with the aim of capturing a weaker notion of dependency---one that is sensitive to data dependencies but insensitive to control dependencies. Viewed through the lens of information flow, this system restricts only  \emph{explicit} flows of information. We make this guarantee precise, and argue why it may be useful for security in practice.

\subsection{Explicit flows}

In their seminal paper on information-flow security, Denning and Denning provided an intriguing characterization of explicit flows \cite{denning1977certification}: ``\emph{\dots an explicit flow [of some information $x$] occurs whenever the operations generating it are independent of the value of $x$.}'' 
Unfortunately, this definition has been largely ignored in the literature. %\footnote{In fact, even in the original paper the distinction between explicit and implicit flows was mentioned only in passing, with no further reference in the rest of the paper.} 
The only related work seems to be Volpano's \cite{volpano99sas}, which defines \emph{weak security} as a trace-based (safety) property: a program is weakly secure if its traces induce secure ``branch-free'' programs. We observe that weak security implies the absence of explicit flow attacks, since information flows in a branch-free program cannot be generated by operations that depend on specific values. (It seems that this connection between Volpano's and Denning and Denning's definitions has not been articulated previously.)

These definitions deserve more attention, since they suggest exactly why explicit flow attacks are so interesting in practice. Explicit flow vulnerabilities are attractive to attackers, since they can be exploited parametrically. Conversely, such vulnerabilities  often point to logical errors rather than implementation ``artifacts", since the information-flow channels are abstract. %Conversely, %because the success of explicit flow attacks cannot actually rely on the specific values involved, 
%For example, if the attacker can write some value to a trusted location, it can write any value to that location instead of worrying how to influence that value. Conversely, 
%if the attacker can read some value from a secret location, it can read any value from that location instead of worrying how to infer that value. 
Finally, various dynamic checks---such as those for exception handling and access control---routinely cause implicit flows in practice. Ignoring these channels not only focuses our attention on other ``definite vulnerabilities'', but also liberates dynamic checks to serve as mechanisms for plugging those vulnerabilities.

%Indeed, if a system is vulnerable to explicit flow attacks, then those attacks are abstract over the specific values being leaked. the success of those attacks does not rely on the specific values being leaked; thus, such attacks can be carried out more predictably than others. 

This may explain why several recent analyses for security have by design ignored implicit flow attacks and focused on eliminating explicit flow attacks~\cite{clause07dytan,shankar06automated,suh04secure,yin07panorama,costa05vigilante,castro06dataflow,vogt07xss,dalton07raksha,broadwell03scrash,chen07format,martin05sql,shankar2001detecting,xie07saturn,zhang02cqual,chaudhuri2009type,tripp2009taj,chaudhuri2009language}. Some of these analyses aim to verify the security of web applications \cite{tripp2009taj,dalton07raksha,shankar2001detecting,vogt07xss}. Many attacks in this context are ultimately due to code injection, and a common defense against such attacks is to sanitize values that may flow from inputs to outputs. %; the sanitization is usually implemented by validating or transforming such values somewhere along the input/output path. Note that 
The sanitization mechanisms merely restrict explicit flows---they may well introduce implicit flows, but such flows are considered benign in this context. Some other analyses aim to formalize security guarantees provided by low-level systems such as file and operating systems \cite{blanchet2008automated,chaudhuri2009type,chaudhuri2009language}, which are usually protected by dynamic access control mechanisms. Preventing explicit flow attacks with these mechanisms already requires some care, and it seems difficult and perhaps undesirable to expect stronger guarantees from such systems. 
%That said, strong information-flow guarantees are necessary in many settings, and it is worthwhile to explore whether such guarantees can be enforced in combination with weaker guarantees as needed, in safe and hopefully useful ways.

\subsection{\dccd}

Our system, \dccd, is a simple variant of \dcc~where the type constructors $T_\ell$ are replaced by $\overline T_\ell$, and the protection mechanisms $\eta_\ell$ are replaced by $\overline \eta_\ell$. These replacements are intended to provide weaker guarantees than their counterparts in \dcc, as discussed above; we enforce them with a slightly different set of rules, which require a new form of type $s^\ell$, called an \emph{open type}. 
Intuitively, the type $s^\ell$ is given to terms of type $s$ that need to be (weakly) protected at level $\ell$. Open types do not have any special introduction or elimination forms. Instead they \emph{qualify} existing types \cite{foster1999theory}, according to the following equations.
\begin{itemize}
\item $(s^\ell)^{\ell'} = s^{\ell \sqcup \ell'}$ and $s = s^\bot$ (protection requirements can be joined with $\sqcup$, and any type can be viewed as an open type with no protection requirement);
\item $\unitT^\ell = \unitT$, $(s \rightarrow t)^\ell = s \rightarrow t^\ell$, $(s \times t)^\ell = (s^\ell \times t^\ell)$, and $\overline T_{\ell'}(s)^\ell = \overline T_{\ell'}(s^\ell)$ (protection requirements are redundant for the unit type, and can be structurally propagated for other non-sum types); 
\item $\overline T_{\ell'}(s)^\ell = \overline T_{\ell'}(s)$ iff $\ell \sqsubseteq \ell'$ (protection requirements can be dropped if there is adequate protection). 
\end{itemize}
%\cenv{Open type equations}{}{
%\vspace{-3mm}
%\[
%\left.
%\begin{array}{rl}
%{\sf (E\mbox{-}open\mbox{-}1)}~ & 
%	{(s^\ell)^{\ell'} = s^{\ell \sqcup \ell'}}
%\\
%{\sf (E\mbox{-}open\mbox{-}2)}~ & 
%	{s = s^\bot}	
%\\
%{\sf (E\mbox{-}unit)}~ & 
%	{\unitT^\ell = \unitT}
%\\ 
%{\sf (E\mbox{-}function)}~ & 
%	{(s \rightarrow t)^\ell = s \rightarrow t^\ell}
%\\ 
%{\sf (E\mbox{-}product)}~ & 
%	{(s \times t)^\ell = (s^\ell \times t^\ell)}
%\\ 
%{\sf (E\mbox{-}effect\mbox{-}1)}~ & 
%	{\overline T_{\ell'}(s)^\ell = \overline T_{\ell'}(s^\ell)}
%\\
%{\sf (E\mbox{-}effect\mbox{-}2)}~ & 
%\inferx
%	{\ell \sqsubseteq \ell'}
%	{\overline T_{\ell'}(s)^\ell = \overline T_{\ell'}(s)}
%\end{array}
%\right.
%\]
%}
%
%These rules may be explained as follows. \erule{open\mbox{-}1} allows such protection requirements to be joined with $\sqcup$, and \erule{open\mbox{-}2} allows any type to be viewed as an open type with no protection requirement. The remaining rules are analogous to \dcc's protection rules. For terms of open unit type $\unitT^\ell$, the protection requirement $\ell$ is redundant---it may be dropped as needed. For terms of open product type $(s \times t)^\ell$, the protection requirement $\ell$ can be propagated to their projections. Similarly, for terms of open function type $(s \rightarrow t)^\ell$, the protection requirement $\ell$ can be propagated to their bodies. Finally, for terms of open protected type $\overline T_{\ell'}(s)^\ell$, the protection requirement $\ell$ can be propagated to their payloads, and can be dropped if $\ell'$ is at least $\ell$.

Note that there is no equation for sum types. In particular, it would be unsafe to equate the open sum type $(s + t)^\ell$ with the sum type $(s^\ell + t^\ell)$, for reasons similar to those discussed in Section \ref{sec:dcc}. It suffices to see that such an equation would imply $(\unitT + \unitT)^\ell = (\unitT^\ell + \unitT^\ell) = (\unitT + \unitT)$.
But recall that the type $(\unitT + \unitT)$ can serve as an encoding of $\mathtt{boolean}$; so the equation in question would allow protection requirements on booleans to be dropped as needed. In general, this would make protection requirements on any data redundant, and completely defeat the purpose of open types. Furthermore, note that by viewing the equations above as rewrite rules from left to right, it is possible to ``normalize" types, effectively pushing the protection requirements that occur in those types as inwards as possible. Such normalization helps maintain syntax-directed types for most terms, except those that have (open) sum types. For the latter terms, we assume that they always have open sum types.

Our enforcement strategy with open types is rather simple. Upon undoing protection of a term of type $\overline T_\ell(s)$, we give it a type $s^\ell$. We then demand that such a term be protected back with a level $\ell'$ that is at least $\ell$. The resulting term has type $\overline T_{\ell'}(s^\ell)$, which can be equated to $\overline T_{\ell'}(s)$, thereby dropping the protection requirement. 
To enforce this strategy, we define (as in \dcc) a predicate $\ell \leq t$, read as ``$t$ is weakly protected at $\ell$", with the following rules: $\ell \leq \unitT$; $\ell \leq (s \rightarrow t)$ iff $\ell \leq t$; $\ell \leq (s \times t)$ iff $\ell \leq s$ and $\ell \leq t$; $\ell \leq (s + t)$ iff $\ell \leq s$ and $\ell \leq t$; and $\ell \leq \overline T_{\ell'}(s)$ iff $\ell \sqsubseteq \ell'$ or $\ell \leq s$. 

Note that there is no rule for open types, since such types are not protected by definition.
On the other hand, we include a rule for sum types. Such a rule is sound in this context because we are only interested in tracking data dependencies and not control dependencies. Indeed, terms of type $(s + t)$---which evaluate to values $\inj i {e}$---cannot leak any data not already leaked by $e$ (which has type either $s$ or $t$). In particular, the constructors $\mathsf{inj}_i$ cannot leak any data, unless the values $\inj i {e}$ already require protection and thus have a non-trivial open type (where the qualifier is not $\bot$)---which is impossible since by the equations above, $(s + t)$ cannot be equal to such a type.

Following \dcc, the typing rules for \dccd~derive judgments of the form $\Gamma; \overline \Pi \vdash e : t$, where $\Gamma$ contains type hypotheses for free variables and $\overline \Pi$ is a (weak) protection context. 
We show only the interesting rules. (The remaining rules are in the appendix.)

\[
\left.
\begin{array}{lc}
{\sf (T^D\mbox{-}case)} & \infer
	{\Gamma; \overline \Pi \vdash e : (s_1 + s_2)^\ell \\ \Gamma, x : s_i^\ell; \overline \Pi \vdash e_i : s}
	{\Gamma; \overline \Pi \vdash \split e x {e_1} {e_2} : s}
\\
\vspace{-3mm} \\
{\sf (T^D\mbox{-}ret)} & \infer
	{\Gamma; \overline \Pi \sqcup \ell \vdash e : s}
	{\Gamma; \overline \Pi \vdash \weak \ell e : \overline T_\ell (s)}
\\ 
\vspace{-3mm} \\
{\sf (T^D\mbox{-}bind)} & \infer
	{\Gamma; \overline \Pi \vdash e : \overline T_\ell (s) \\ \Gamma, x : s^\ell; \overline \Pi \vdash e' : t \\ \ell \leq \overline T_{\overline \Pi}(t)}
	{\Gamma; \overline \Pi \vdash \bind x e {e'} : t}
\end{array}
\right.
\]

${\sf (T^D\mbox{-}case)}$ assumes that the case construction $\inj i e$ has an open sum type, and propagates its protection requirement to the variable $x$ bound to $e$ at run time. This allows sensitive data to be safely destructed, without losing track of its protection requirements. 
All other rules are syntax-directed (thanks to normalization of types as mentioned above) and are analogous to those in \dcc. 
In particular, ${\sf (T^D\mbox{-}ret)}$ states that $\weak \ell e$ has type $\overline T_\ell(s)$ whenever $e$ has type $s$, assuming (weak) $\ell$-protection by the context. ${\sf (T^D\mbox{-}bind)}$ states that $(\bind x e {e'})$ has type $t$ only if $e$ has a type of the form $\overline T_\ell(s)$, $e'$ has type $t$ assuming that $x$ has open type $s^\ell$, and the type $\overline T_{\overline \Pi}(t)$ is (weakly) protected at $\ell$. The latter condition means that either $t$ is protected at $\ell$, or $\overline \Pi$ is at least $\ell$. This ensures that the result of $e'$ cannot leak any data at $\ell$, including any data in $x$, which is bound to the result of $e$ upon undoing its $\ell$-protection at run time.

We formalize the key property of this type system using a type-directed \emph{safety} relation over terms, $e~\rhd_\ell : s$, meaning that term $e$ of type $s$ is safe at level $\ell$. Our safety relation relies on a semantics with ``taint propagation''. Thus, we extend the internal syntax with terms of the form $e^\ell$, meaning $e$ tainted with $\ell$---intuitively, $e^\ell$ is similar to $(\bind x {\weak \ell e} x)$ for fresh $x$. We define equations over tainted terms, closely following the equations over open types. Thus we have: $(e^\ell)^{\ell'} = e^{\ell \sqcup \ell'}$; $e = e^\bot$; $\unit^\ell = \unit$; $(\abs x e)^\ell = \abs x {e^\ell}$; $\tup {e_1} {e_2}^\ell = \tup {e_1^\ell} {e_2^\ell}$; $\weak {\ell'} e^\ell = \weak {\ell'} {e^\ell}$; and $\weak {\ell'} e^\ell = \weak {\ell'} {e}$ iff $\ell \sqsubseteq \ell'$.
As usual, these equations let us normalize terms so that only terms of sum types carry taints. Finally, we %include a global reduction rule---if $e$ reduces to $e'$ then $e^\ell$ reduces to $e'^\ell$---and 
extend the local reduction rules for $\mathsf{bind}$ and $\mathsf{case}$ as follows: $(\bind x {\weak \ell e} {e'})$ reduces to $e'[e^\ell/x]$, and $(\split {{\inj i e}^\ell} x {e_1} {e_2})$ reduces to $e_i[e^\ell/x]$. 
Thus, we taint a term upon undoing its protection, and propagate the taint on a term to its subterm upon pattern matching. Note that such taint propagation ignores implicit flows. We use this semantics in the derivation rules of our safety relation, as follows.
\begin{itemize}
\item $e ~\rhd_\ell : s$ iff $e$ reduces to $v$ and $v~\rhd_\ell : s$
\item $\unit~\rhd_\ell : \unitT$
\item $v~\rhd_\ell : (s \rightarrow t)$ iff for all $e$, if $e ~\rhd_\ell : s$ then $\app v e ~\rhd_\ell : t$
\item $\tup {e_1} {e_2}~\rhd_\ell : (s_1 \times s_2)$ iff $e_1 ~\rhd_\ell : s_1$ and $e_2 ~\rhd_\ell : s_2$
\item $\inj i {e_i}~\rhd_\ell : (s_1 + s_2)$ iff $e_i ~\rhd_\ell : s_i$
\item $\weak {\ell'} e~\rhd_\ell : \overline T_{\ell'}(s)$ iff $\ell' \not\sqsubseteq \ell$ or $e ~\rhd_\ell : s$
\end{itemize}

Thus, safety is analogous to indistinguishability as defined in Section \ref{sec:dcc}, except that we are concerned with properties of a single term rather than a pair of terms. As expected, tainted terms are unsafe, and $\weak {\ell'} e$ is safe at $\ell$ unless $\ell$ is at least $\ell'$. %{\sf (U-open)} allows protection requirements to be erased, since they have no run-time significance. 

For example, let $e = \weak \ell {\inj i \unit}$ for some $i \in \{1,2\}$ and suppose that $\ell \not\sqsubseteq \ell'$. Then $e~\rhd_{\ell'} : \overline T_\ell(\unitT + \unitT)$. Now recall the functions $f$ and $g$ defined in Section \ref{sec:intro} (and assume that $s_1 = s_2 = \unitT$ for simplicity). Clearly $f {\not\rhd}_{\ell'}$, since $\app f e$ reduces to ${\inj i \unit}^\ell$ and $\inj i \unit %\linebreak[4]
{\not\rhd}_{\ell'}$. Fortunately, $f$ is not typable in \dccd. In contrast, we can show that $g~\rhd_{\ell'} : \overline T_\ell(\unitT + \unitT) \rightarrow (\unitT + \unitT)$, and $g$ is typable in \dccd. Next consider $f' = \abs x {\weak \ell {\app f x}}$. It is easy to check that $f'~\rhd_{\ell'} : \overline T_\ell(\unitT + \unitT) \rightarrow \overline T_\ell(\unitT + \unitT)$, and $f'$ is typable in \dccd.
Indeed, the type system guarantees that whenever a typed function is applied to (weakly) $\ell$-protected inputs, it always produces outputs that are safe at levels that are not at least $\ell$. 

\begin{theorem}[\dccd~soundness]\label{thm:dccd}
If $~\vdash e : \overline T_\ell(s) \rightarrow t$ and $~\vdash e' : s$, then for any $\ell'$ such that $\ell \not\sqsubseteq \ell'$, $\app e {\weak \ell {e'}} ~\rhd_{\ell'} : t$. 
\end{theorem}
Furthermore, we show that \dccd's type system is at least as liberal than \dcc's, by defining an appropriate encoding between the two systems. (In fact, it is strictly more liberal by the example above.)

\begin{theorem}[\dcc~to \dccd]\label{thm:dccd-dcc} Let $[\![ \cdot ]\!]$ translate terms and types by replacing $\strong \ell \cdot$ with $\weak \ell \cdot$, and $T_\ell(\cdot)$ with $\overline T_\ell(\cdot)$. If $\vdash e : s$ in \dcc~then $\vdash [\![ e ]\!] : [\![ s ]\!]$ in \dccd. 
\end{theorem}

\subsection{Remarks}

Before we move on, let us try to carefully understand the guarantee provided by \dccd.

\dccd's semantics, based on taint propagation, is closely related to Volpano's execution monitor for weak security \cite{volpano99sas}. In fact, results of evaluation in \dccd~can be interpreted as branch-free \dcc~programs ``induced by traces'', and typing in \dccd~guarantees security of such programs in \dcc. 
\begin{theorem}[\dccd~soundness, \`a la Volpano \cite{volpano99sas}]\label{thm:dccd-volpano}
Let $\{\!\!\{\cdot\}\!\!\}$ translate terms and types by replacing $\weak \ell \cdot$ with $\strong \ell \cdot$, $\overline T_\ell(\cdot)$ with $T_\ell(\cdot)$, and $(\cdot^\ell)$ with $(\bind x {\strong \ell \cdot} x)$ for fresh $x$. If $~\vdash e : \overline T_\ell(s) \rightarrow t$, $~\vdash e' : s$, and $\app e {\weak \ell {e'}}$ evaluates to $v$ in \dccd, and if no protection type occurs negatively in $t$, then $\vdash \{\!\!\{v\}\!\!\}: \{\!\!\{t\}\!\!\}$ in \dcc. 
\end{theorem}
For example, consider the following function of type $\overline T_\ell(\unitT + s) \rightarrow (\unitT + \overline T_\ell(s))$:
$$k = \abs x {\bind y x {\split y z {\inj1 \unit} {\inj2 {\weak \ell z}}}}$$
Let $e$ be any term of type $s$; we have that $\app k {\weak \ell {\inj2 e}}$ reduces to $\inj2 {\weak \ell {e^\ell}}$, which translates via $\{\!\!\{\cdot\}\!\!\}$ to $\inj2 {\strong \ell {\bind w {\strong \ell e} w}}$. The latter is a branch-free, typed, \dcc~program. In fact, by the theorem above, all branch-free programs induced by traces of $k$ are typed, and thus $k$ is weakly secure. In contrast, if the protection $\weak \ell \cdot$ in the body of $k$ is dropped, the induced branch-free program does not remain typable.

Furthermore,  \dcc's type system eliminates explicit flow attacks as characterized by Denning and Denning \cite{denning1977certification}, since we have already argued that weak security implies the absence of such attacks.
Note that an explicit flow attack can be camouflaged as an
implicit flow attack by ``deep copying'', \emph{i.e.}, by destructing a sensitive term all the way down with elimination
forms and constructing it back from scratch with introduction forms. Formally, let $\mathsf{erase}$ be a function on types that erases the label qualifiers in open types.
Thus, for any $s$, we have $\ell \leq \mathsf{erase}(s)$ for all $\ell$; in other words, the side condition in $({\sf T^D\mbox{-}bind})$ is redundant for erased types.
Now we can define a family of functions $\emph{leak}_\ell(t) : \overline T_\ell( t ) \rightarrow \mathsf{erase}(t)$ that 
behave just like $\abs {x: \overline T_\ell(t)} \bind y x y$, such that the former are typable in \dccd, but the latter are not (see the appendix).
Thus, in the limit we may be assured nothing even if \dccd~deems our program ``secure''---while \dccd~guarantees that all explicit leaks are eliminated, these leaks may remain hidden in the guise of implicit leaks (which remain unrestricted). However, we argue that \dccd~still provides ``pretty good protection'', at least for code that the attacker cannot fully control. 
Indeed, for such non-malicious code, we may assume that the programmer does not try to intentionally circumvent our analysis. Under this assumption, prioritizing explicit flows over implicit flows is arguably reasonable, for several reasons: 
\begin{itemize}
\item %If a programmer intends to transfer some value between variables, he'll copy it directly! Ê
No sane programmer would copy all bits of some value indirectly, one at a time, instead of copying the value directly. 
\item As argued in \cite{russo09implicit}, implicit leaks are largely harmless for non-malicious code, since such leaks cannot be exploited efficiently by the attacker.
\item As shown in \cite{king2008implicit}, checking for implicit flows can be costly to the programmer---typically lots of false alarms arise in systems that check for implicit flows.
\end{itemize}

%
% ÊSince implicit leaks are rare in practice, such a system provides "pretty good protection." ÊIt might even be interesting to include the reviewer's transformation in your paper, to show how all of a datum can be leaked. ÊThus you show you are not concerned by this fact----you still want to understand what programs that use only explicit flow tracking give you compared to implicit+explicit.

%
%The most negative reviewer rightly points out that you can always transform a program to leak a value indirectly, rather than directly. ÊBut I think we need to think of the "attacker model" here: who is it that is hypothetically transforming my program? Ê

%That is, who wrote the program I'm analyzing for leaks? Ê

%

%On closer inspection, 

\section{Dynamic weakening in \dccdc}\label{sec:dccdc}

While \dccd-style protection is sufficient in some settings, \dcc~still enjoys better theoretical foundations and promises many desirable properties that \dccd~cannot. In practice, we should be able to mix \dccd-style protection carefully with \dcc-style protection as needed, and still be able to reason precisely about the guarantees of the resulting systems, short of weakening all the guarantees provided by \dcc-style protection. 
We investigate these issues in the setting of a hybrid language \dccdc.

\subsection{\dccdc}

\dccdc's syntax and typing rules are obtained by merging those of  \dcc~and \dccd. The merge is mostly straightforward; we make a few adjustments to encourage the two subsystems to interact. (The full system is available for reference in the appendix.) First, we carry both kinds of protection contexts in typing judgments, and modify the \dcc~rule ${\sf (T\mbox{-}ret)}$ as follows. 
$$\infer
	{\Gamma; \Pi \sqcup \ell ; \overline \Pi \sqcup \ell \vdash e : s}
	{\Gamma; \Pi; \overline \Pi \vdash \strong \ell e : T_\ell (s)}
$$
Thus, any \dcc-style protection provided by the context is made evident not only in its usual protection context, but also in the weak protection context. 
Next, we add the following protection rules and open type equations: $\ell \preceq \overline T_{\ell'}(s)$ if $\ell \preceq s$; $\ell \leq T_{\ell'}(s)$ if $\ell \leq \overline T_{\ell'}(s)$; and $T_{\ell'}(s)^\ell =  T_{\ell'}(s^\ell)$ if $\overline T_{\ell'}(s)^\ell =  \overline T_{\ell'}(s^\ell)$.
%These rules can be explained as follows. On the one hand, ${\sf (P\mbox{-}effect)}$ conservatively assumes that terms of type $\overline T_{\ell'}(s)$---which evaluate to weakly protected terms---can leak just as much information as their payloads, which are of type $s$. On the other hand, ${\sf (P^D\mbox{-}monad)}$ considers the type $T_{\ell'}(s)$ to be weakly protected whenever the weaker type $\overline T_{\ell'}(s)$ is so protected; 
These rules internalize the fact that \dcc's protection types subsume \dccd's protection types, as shown in Theorem~\ref{thm:dccd-dcc}. 
%${\sf (E\mbox{-}monad)}$ is based on similar reasoning. 
In particular, these rules admit functions such as $\abs x {\bind y x {\strong \ell y}}$ of type $\overline T_\ell(s) \rightarrow T_\ell(s)$, that can be used to \emph{strengthen} protection on terms.
Finally, we unify the rules for non-protection types; in particular we have:
$${\sf (T^{DC}\mbox{-}case)} ~~ \infer
	{\Gamma; \Pi; \overline \Pi \vdash e : (s_1 + s_2)^\ell \\ \Gamma, x : s_i^\ell; \Pi; \overline \Pi \vdash e_i : s}
	{\Gamma; \Pi; \overline \Pi \vdash \split e x {e_1} {e_2} : s}
$$
%
%is based on ${\sf (T^D\mbox{-}case)}$, since it is more general than ${\sf (T\mbox{-}case)}$ via ${\sf (E\mbox{-}open\mbox{-}2)}$. 
% 
The definitions of indistinguishability and safety are similarly extended, and we can show that the respective guarantees of \dcc~and \dccd~are preserved in \dccdc. %Of course, this requires adding the appropriate rules to the definitions of the indistinguishability and safety relations, as follows.
%\cenv{Additional rules for indistinguishability and safety relations}{}{
%\vspace{-2mm}
%%
%\[
%\left.
%\begin{array}{lc}
%{\sf (I\mbox{-}effect)} \quad & \infer
%	{e \sim_\ell e' : s}
%	{\weak {\ell'} e \sim_\ell \weak {\ell'} {e'} : \overline T_{\ell'}(s)}
%\\
%\vspace{-3mm} \\
%%{\sf (I\mbox{-}open)} \quad & \infer
%%	{v \sim_\ell v' : s}
%%	{v \sim_\ell v' : s^\ell}
%%\\
%{\sf (S\mbox{-}monad)} \quad & %\!\!\!\!\!\!\!\!\!\!\!\!\!\!\!\!\!\!\!\!\!\!\!\!\!\!\!\!\!\!\!\!\!\!\!\!\!
%\infer
%	{\weak {\ell'} e~\rhd_\ell : \overline T_{\ell'}(s)}
%	{\strong {\ell'} e~\rhd_\ell : T_{\ell'}(s)}
%\end{array}
%\right.
%\]
%}
%
\begin{theorem}[\dccdc~soundness, preliminary]\label{thm:dccdc-plain} Theorems \ref{thm:dcc} and \ref{thm:dccd} also hold in \dccdc.
\end{theorem}

\vspace{-3mm}
\subsection{A weakening primitive}

Next we include a $\mathsf{weaken}$ primitive in \dccdc, which acts as a further bridge between the two subsystems (going in the opposite direction as the strengthening functions above). Our intention is that such a primitive should allow terms of type $T_\ell(s)$ to be viewed as terms of type $\overline T_\ell(s)$, possibly with some caution.

Unsurprisingly, using $\mathsf{weaken}$ may invalidate the protection guarantees provided by \dcc's types. As a simple example, consider the following function:
\begin{eqnarray*}
h & = & \abs x {\bind y {\weaken x} {\split y z {\inj 1 \unit} {\inj 2 \unit}}}
\end{eqnarray*}
Assuming a typing rule that allows $\weaken e$ to have type $\overline T_\ell(s)$ whenever $e$ has type $T_\ell(s)$, this function can be typed $T_\ell(\unitT + \unitT) \rightarrow (\unitT + \unitT)$. However, $h$ clearly has an information flow violation; formally, we have that $\app h {\strong \ell {\inj 1 \unit}} \not\sim_\ell \app h {\strong \ell {\inj 2 \unit}}$, which contradicts Theorem \ref{thm:dccdc-plain}. Worse, $h$ can be used as an oracle to generate more complex counterexamples. Consider the following functions:
\vspace{-1mm}
\begin{eqnarray*}
m & = & \abs x {\strong \ell {(\bind y x {\split y z {\inj 1 \unit} {\inj 2 \unit}})}} \\
n & = & \abs x {\app h {\app m x}}
\end{eqnarray*}
The function $m$ can be typed $T_{\ell'}(s + t) \rightarrow T_\ell(\unitT + \unitT)$ in \dcc~as long as $\ell' \sqsubseteq \ell$, and does not leak information on $x$ per se; it derives a bit of information on $x$ and protects that bit before returning it. Still, the function $n$ with type $T_{\ell'}(s + t) \rightarrow (\unitT + \unitT)$ is able to use $m$ in combination with $h$ to leak that bit.

As this example suggests, using $\mathsf{weaken}$ at level $\ell$ in a program may invalidate \dcc-style guarantees for all types protected by levels $\ell$ and lower. However, weaker \dccd-style guarantees should still hold for such types (because there is no way to get around \dccd's typing rules). Moreover, assuming that there are no other uses of $\mathsf{weaken}$ in the program, we expect that stronger \dcc-style guarantees should remain valid for all other types. The reason is that such types, which are protected by levels higher or incomparable to $\ell$, will never delegate the responsibility of protection to the weakened types. In summary, we can precisely reason about protection in this system as long as we carefully track the uses of $\mathsf{weaken}$ in the program. 

Curiously enough, such an analysis can be viewed as a special case of \dcc's dependency analysis, just like many other applications of \dcc. Indeed, the original motivation for studying \dcc~was its ability to express various program analyses---including call tracking, slicing, partial evaluation, as well as information-flow control---in a uniform setting. Our analysis is similar in spirit, and can be expressed by recycling \dcc's types to carry \emph{blames} for weakening. 

Specifically, we consider a lattice of blames that is isomorphic to the lattice of levels, \emph{i.e.}, for each level $\ell$ we have a blame $\beta(\ell)$, where $\beta$ is some lattice isomorphism. Then, instead of the na\"ive typing rule for $\mathsf{weaken}$ above, we include the following rule: \vspace{-1mm}
$${\sf (T^{DC}\mbox{-}weaken)} \quad \infer
	{\Gamma; \Pi; \overline \Pi \vdash e : T_\ell (s)}
	{\Gamma; \Pi; \overline \Pi \vdash \weaken e : T_{\beta(\ell)} (\overline T_\ell (s))}
$$
Intuitively, this means that whenever we use $\mathsf{weaken}$ to view terms of type $T_\ell(s)$ as terms of type $\overline T_\ell(s)$ in a program, we simultaneously blame $\beta(\ell)$ for facilitating such a view. While this allows us to get away with weaker protection requirements on such terms, it also forces some caution: the blame must be carried around whenever a result depends on those terms. Fortunately, \dcc's typing rules can enforce this for free.

A reassuring interpretation of blames may be obtained through the lens of the Curry-Howard isomorphism, following a recent reading of \dcc~as an authorization logic~\cite{abadi2007access}. Specifically, we can interpret the blame $\beta(\ell)$ as a principal that controls protection requirements at level $\ell$, and rewrite the type of $\weaken e$ as $\beta(\ell)~\mathsf{says}~(\overline T_\ell (s))$. 
Using the logic, we can now pinpoint the principals whose statements may have influenced protection requirements in a program, resting assured that the protection guarantees at other levels will not be influenced by these statements.

\subsection{Blame orderings}

Note that we have not yet specified how the ordering in the blame lattice should be related to $\sqsubseteq$. One interesting scenario is where the ordering is the same, so that $\beta$ preserves joins and meets. In this scenario, the type of a program must carry a blame $\beta(\ell)$ such that $\ell$ upper-bounds the levels of weakening on which its results may depend. (This is because \dcc's rules guarantee that $\beta(\ell)$ will upper-bound the levels of weakening on which the results of the program may depend.) In other words, \dcc-style protection guarantees must hold at all levels not $\ell$ or lower. 

Formally, we define the blame $\mathcal B(t)$ carried by a program of type $t$ as the join of all blames that appear in $t$. We then prove the following theorem. %Instead, we use the fact that if $\beta(\ell) \not\leq t$ then $\beta(\ell) \not\sqsubseteq \mathcal B(t)$, and show the following theorem.
\begin{theorem}[\dccdc~soundness: strong protection]\label{thm:dccdc}
If $~\vdash e : T_\ell(s) \rightarrow t$, $~\vdash e_1 : s$, $~\vdash e_2 : s$, and $\ell$ is any label such that $\ell \not\sqsubseteq \beta^{-1}(\mathcal B(t))$, then for any $\ell'$ such that $\ell \not\sqsubseteq \ell'$, $(e~(\eta_\ell~e_1)) \sim_{\ell'} (e~(\eta_\ell~e_2)) : t$. Moreover, Theorem \ref{thm:dccd} holds as is in this system.
\end{theorem}
As a simple example, consider the following well-typed program of type $T_{\beta(\ell)}(\unitT + \unitT)$ (where $i \in \{1,2\}$):
$$\bind x {\weaken {\strong \ell {\inj i \unit}}} {\strong {\beta(\ell)} x}$$
We have $\mathcal B(T_{\beta(\ell)}(\unitT + \unitT)) = \beta(\ell)$, so we can be sure that this program does not (and cannot be used to) weaken \dcc-style protection guarantees at $\ell'$ unless $\ell' \sqsubseteq \ell$. 

An equally interesting scenario is where we flip the ordering in the blame lattice, so that $\beta$ exchanges joins and meets. In this scenario, the type of a program must carry a blame $\beta(\ell)$ such that $\ell$ lower-bounds the levels of weakening on which its results may depend. (Again, this is because \dcc's rules guarantee that $\beta(\ell)$ will upper-bound the levels of weakening on which the results of the program may depend.) In other words, \dccd-style protection guarantees must be robust against all levels not $\ell$ or higher.
%In other words, $\ell$ provides a measure of robustness for these guarantees.

Formally we prove the following theorem, where $\mathcal B(t)$ is defined as earlier.
\begin{theorem}[\dccdc~soundness: weak protection]\label{thm:dccdc}
Suppose that $~\vdash e : T_\ell(s)$, $~\vdash e' : \overline T_\ell(s) \rightarrow t$,  and $\ell$ is any label such that $\beta^{-1}(\mathcal B(t)) \not\sqsubseteq \ell$. Then it is impossible to derive $~\vdash \app {e'} {(\bind x {\weaken e} x)} : t$.
\end{theorem}
Continuing the previous example, we can be sure that \dccd-style guarantees for the program are not influenced by weakening at level $\ell'$ unless $\ell \sqsubseteq \ell'$.

\section{Precise dependency analysis in \dcccd}\label{sec:dcccd}

%If we had to make a movie out of the story so far, we would likely not win any awards. We have a kingdom run by a blue-blooded tyrant, \dcc; a rustic rebel, \dccd, conquers the kingdom and ushers in some liberalization, but it destabilizes the kingdom so much that the tyrant has to be rushed back in disguise to restore order! Indeed, what this story misses is a climax: can the rebel make a comeback, showing the tyrant that such liberalization will in fact help run his kingdom better?

Just as a \dcc-style analysis can strengthen protection guarantees in a hybrid system, in turns out that a \dccd-style analysis can improve the coverage of such guarantees. In this section, we deconstruct information flow control in \dcc~into two separate problems: one of restricting explicit flows, and the other of restricting implicit flows. The former is already handled by \dccd; the latter, which is entirely due to case analysis, can be handled by reworking some of the rules for sum types in \dcc. The resulting system, \dcccd, becomes more liberal than \dcc~without compromising its guarantees. We discuss the benefits of such an enhancement towards the end of the section.

\subsection{\dcccd}

\dcc~conservatively assumes that case constructors may always convey sensitive information; thus, it restricts both explicit and implicit flows in one shot by requiring that sum types can never be considered protected (see the discussion on \dcc's protection rules in Section \ref{sec:dcc}). 
Unfortunately this restriction causes several benign programs to be rejected by \dcc~simply because they use case construction.
We relax this restriction by observing that any information leak is ultimately due to either an explicit leak through data flow or an implicit leak through control flow. Specifically, evaluating a term of sum type may reveal information about sensitive data only if that term either does a case analysis on sensitive data, or releases the sensitive data itself. 

Technically, this separation of concerns is already somewhat evident in \dccd, where we weaken \dcc's protection rules to allow sum types to be considered protected (see Section \ref{sec:dccd}). But by itself this is unsound, given the dangerous nature of case constructors---it admits both explicit and implicit flows. Thus, we also require open types---types with qualifiers to precisely track data flow through programs---and we use these qualifiers to restrict explicit flows in \dccd. 
%Thus, all we had to do was make a small change in the rule for case analysis to accommodate terms with qualified sum types.   
In particular, the typing rule for case analysis needs to accommodate terms with qualified sum types, because the qualifiers can be eliminated on all types other than sum types---they ``stick" to sum types exactly because of the dangerous nature of case constructors. 
While in \dccd~we choose to ignore implicit flows caused by such case analysis, in \dcccd~we do not. 
%In the system we study below, the same qualifiers also help restrict implicit flows. 

Note that in order to adjust the rule for case analysis to account for implicit flows, we must have some idea of the level of information that we are interested in protecting---otherwise, we would have to conservatively ban any case analysis. For this purpose, we need to carry an \emph{open context} $\Sigma$ in typing judgments, which indicates the minimum level of protection required by the context. For closed terms, $\Sigma$ is $\top$. 

The developments of Section \ref{sec:dccdc} are orthogonal to our present purposes, so we drop terms of the form $\weaken e$ and $\weak \ell e$ in the language; indeed, on the surface we do not care about \dccd-style protection at all, although \dccd's type system is an important component of the system internally. Accordingly, we also drop weak protection contexts. The remaining system mostly inherits from \dccdc; we make a few adjustments, discussed below. (The full system is available for reference in the appendix.) 

We now have two typing rules for $\mathsf{bind}$, both offering \dcc-style protection. The first rule is similar to that in \dcc. 
$$
{\sf (T^{CD}\mbox{-}bind\mbox{-}old)} \quad \infer*[sep=5mm]
	{\Gamma; \Pi; \Sigma \vdash e : T_\ell (s) \\ \Gamma, x : s; \Pi; \Sigma \vdash e' : t \\ \ell \preceq T_\Pi(t)}
	{\Gamma; \Pi; \Sigma \vdash \bind x {e} {e'} : t}
$$
The other rule is new, and captures the interaction of the two subsystems.
$${\sf (T^{CD}\mbox{-}bind\mbox{-}new)} \quad \infer*[sep=5mm]
	{\Gamma; \Pi; \Sigma \vdash e : T_\ell (s) \\ \Gamma, x : s^\ell; \Pi; \Sigma \sqcap \ell \vdash e' : t \\ \ell \leq T_\Pi(t)}
	{\Gamma; \Pi; \Sigma \vdash \bind x {e} {e'} : t}
$$
Curiously, this rule looks similar to ${\sf (T^{D}\mbox{-}bind)}$ in \dccd, although functionally it is intended to be closer to ${\sf (T\mbox{-}bind)}$ in \dcc. %The similarities are as important as the differences. 
Like ${\sf (T\mbox{-}bind)}$, ${\sf (T^{CD}\mbox{-}bind\mbox{-}new)}$ applies to terms of type $T_\ell(s)$ instead of $\overline T_\ell(s)$. On the other hand, like ${\sf (T^{D}\mbox{-}bind)}$, we use the weak protection predicate $\leq$ instead of $\preceq$, while assuming an open type for $x$. This takes care of explicit flows, but not implicit flows. In addition, to handle implicit flows, we meet $\ell$ with the open context, deferring their actual restriction till we encounter case analysis at level $\ell$.

The new rule for case analysis is as follows.
$$
{\sf (T^{CD}\mbox{-}case)} \quad \infer*[sep=5mm]
	{\Gamma; \Pi; \Sigma \vdash e : (s_1 + s_2)^\ell \\  \Sigma \not\sqsubseteq \ell \\ \Gamma, x : s_i^\ell; \Pi; \Sigma \vdash e_i : s}
	{\Gamma; \Pi; \Sigma \vdash \split e x {e_1} {e_2} : s}
$$
As in ${\sf (T^D\mbox{-}case)}$, this rule requires---without loss of generality---that $e$ have an open sum type, with some protection requirement $\ell$. In addition, it requires that the open context $\Sigma$ be no lower than $\ell$---so that any implicit flows at $\ell$ that may occur through the case analysis are irrelevant to (\emph{i.e.}, cannot compromise) $\Sigma$. 
%Note that this requirement does not apply if $\ell$ is $\bot$. Indeed, $\ell$ may be $\bot$ for various reasons, and for none of those reasons is this requirement necessary: 
%\begin{itemize}
%\item either $e$ flows from a $\bot$-protected term, in which scenario we do not care about protecting $e$; 
%\item or $e$ actually has type $(s_1 + s_2)$, which means that:
%\begin{itemize} 
%\item either it does not flow from a protected term; 
%\item or its protection requirement was omitted because ${\sf (T^{CD}\mbox{-}bind\mbox{-}1)}$ was applied instead of ${\sf (T^{CD}\mbox{-}bind\mbox{-}2)}$, in which scenario the side condition invoking \dcc's protection rules is already sufficient to rule out implicit flows. 
%\end{itemize}
%\end{itemize}
%
With these rules, we show that \dcccd~provides the same guarantees as \dcc, and is at least as liberal.
\begin{theorem}[\dcccd~soundness and completeness]\label{thm:dcccd}
If $~\vdash e : T_\ell(s) \rightarrow t$, $~\vdash e_1 : s$, and $~\vdash e_2 : s$, then for any $\ell'$ such that $\ell \not\sqsubseteq \ell'$, $\app e {\weak \ell {e_1}} \sim_{\ell'} \app e {\weak \ell {e_2}} : t$. Furthermore, if $\vdash e' : s'$ in \dcc~then $\vdash e' : s'$ in \dcccd.
\end{theorem}
In fact, \dcccd~accepts more programs than \dcc. For example, the following functions---rejected by \dcc~(see Section \ref{sec:intro})---have type $T_\ell(s) \rightarrow (\unitT + \unitT)$ in \dcccd:
$$\abs x {\bind y x {\inj i \unit}} \qquad i \in \{1,2\}$$
%Likewise, the following function---rejected by \dcc---has type $T_\ell(s) \rightarrow (s_1 + s_2) \rightarrow (\unitT + \unitT)$ in \dcccd:
%\begin{eqnarray*}
%&&\abs x {\abs w {\bind y x \\
%&& \qquad {\split w z {\inj 1 \unit} {\inj 2 \unit}}}}
%\end{eqnarray*}
%%
%One might argue that these programs are not all that interesting, since the problematic parts of the code are dead; indeed, \dcc~is intended to be an intermediate language for compilation (rather than a source language), and such dead code can be translated out before typing. However, there are other non-trivial examples for which there may not be a general mechanism to translate out typing problems.
%
As a more interesting example, consider the function $\emph{switch}$ below---rejected by \dcc---which has type $T_\ell(\mathtt{boolean}) \rightarrow \mathtt{boolean} \rightarrow \mathtt{option}(T_\ell (\mathsf{boolean}))$ in \dcccd. (We use the encodings $\mathtt{boolean} = (\unitT + \unitT)$, $\mathtt{option}(\alpha) = (\unitT + \alpha)$, $\mathtt{false} = \inj1 \unit$, $\mathtt{true} = \inj2 \unit$, 
$(\mathtt{if}~e~\mathtt{then}~e_2~\mathtt{else}~e_1) = (\split e \_ {e_1} {e_2})$, $\mathtt{none} = \inj1 \unit$, and $(\mathtt{some}~e) = \inj2 e$.
\begin{eqnarray*}
\emph{switch} & = & \abs x {\abs b \bind {b'} x {\app {\app {\emph{match}} b} {b'}}} \\
\emph{match} &  =  &  \abs b {\abs {b'} \mathtt{if}~b~\mathtt{then}~\mathtt{none}~\mathtt{else}~(\mathtt{some}~{\strong \ell {\app {\emph{not}} {b'}}})} \\
\emph{not} & = & \abs {b'} {\mathtt{if}~b'~\mathtt{then}~\mathtt{false}~\mathtt{else}~\mathtt{true}} 
\end{eqnarray*}
In general, undoing protection of terms early in the control-flow graph seems to cause problems in \dcc, but not in \dcccd.
\subsection{Remarks}

One may, of course, wonder whether our enhancement of \dcc's type system is at all necessary. Indeed, \dcc~is designed to be a target language in which (type-based) program analyses can be encoded to prove their soundness: typing derivations in the source language are translated to typing derivations in DCC, and the soundness of the latter is used to reason about the soundness of the former. In this sense, in fact it is possible to encode \dcccd~in \dcc: we compile \dcccd~programs to the SLam calculus \cite{heintze98slam} by erasing $\mathsf{bind}$s, and then use the well-known encoding of SLam in \dcc~\cite{abadi1999core}. Thus, \dcc's status as a core calculus of dependency is not challenged.
However, as in most such encodings, the translated \dcc~programs are not syntactically equivalent to the source programs. In particular, $\mathsf{bind}$s may be pushed inwards and duplicated across branches. Reasoning about the soundness of this translation requires exactly those observations that underlie the design of \dcccd. %, which thus says something about the soundness of the target; reasoning with semantics, we can then hope to say something about the soundness of the source
Furthermore, the translated programs are inefficient. Indeed, in implementations of \dcc~in the polymorphic lambda calculus \cite{tse2004translating}, $\mathsf{bind}$s are implemented as applications of secret keys (decryptions) to protection abstractions (encryptions)---and it makes sense to pull such applications as outwards as possible for efficiency. For source languages with \dcc-like primitives, it is reasonable to expect that programs will be already be optimized; and we have shown that \dcc-like typing rules do not preserve typability for such optimizations. In summary, we believe that deconstructing information-flow analysis into explicit-flow and implicit-flow analysis, as in \dcccd, provides a better guideline for designing type systems for \dcc-like source languages, than placing an overall restriction on sum types, as in \dcc. 
%Typability should be preserved for such optimizations.
%In general, sound optimizations may break typing (so the soundness of such optimizations must be proved separately). But if such a proof is available, surely a target program can be typed and then optimized, so why is it sensible to enhance completeness for a target language? Because the source program may be already optimized! 
Other enhancements along such lines have been suggested previously \cite{tse2004translating}.

\section{Discussion}\label{sec:relwork}
For brevity, in this paper we have omitted any discussion of pointed types and recursive programs, although they do appear in \dcc~\cite{abadi1999core}. However, we have checked that including these elements does not cause any problems in our results---which is hardly surprising since nontermination does not play an interesting role for weak security.

We have tried to remain close in spirit to Volpano's definition of weak security and Denning and Denning's characterization of explicit flows in our formal definition of \dccd. However, inherent differences in the underlying languages make it difficult to establish a formal correspondence.

%In particular, if an insecure information flow is caused by a combination of ``implicit" and ``explicit" leaks, we prefer to consider such a flow explicit simply because we want \dccd~to restrict such a flow. Indeed, we label only those flows implicit that can be eliminated by moving to a nondeterministic semantics for case analysis. Whether this classification ``agrees" with Denning and Denning's is, in the end, irrelevant to security. 

%Finally, it would seem that in a practical implementation of \dccdc~the overhead of carrying blames would be too painful. However, note that it should not be too difficult to infer the blames required to type a program; and conversely, once a program is typed, the blames can be erased. 

There is a huge body of research on noninterference-based security for languages; see \cite{sabelfeld2003language} for a survey. However, there seems to be a disconnect between this research and most security tools implemented in practice, which ignore implicit flows. Some interesting previous studies have tried to explain why, and under what circumstances, it may make sense to ignore implicit flows in practical security \cite{king2008implicit,russo09implicit}. Unfortunately, we do not know of any work on formalizing the resulting safety guarantees of such tools, although \cite{volpano99sas} provides some valuable insights and several security type systems for process calculi have been designed around similar ideas~\cite{abadi02sectyplog,cardelli2005secrecy,chaudhuri2009type,chaudhuri2009language}. 

The idea of mixing strong and weak dependency analysis in mutually benefitial ways appears to be new. Indeed, our results suggest some interesting ways in which noninterference-based security may be reconciled with trace-based security within the same system, enhancing soundness of the latter and completeness of the former. Specifically, in a system where protection may have been partially weakened, a strong blame analysis can be used to provide strong protection guarantees for those parts of the system that are not affected by such weakening. Conversely, a weak flow analysis can be used to increase the coverage of such guarantees. 

We hope that these results will spur further interest in bridging the gap between these two views of security.

%\paragraph{Acknowledgments}
%This work benefitted from discussions with Steve Zdancewic, Michael Hicks, Aslan Askarov, and David Mazi\`eres.

\bibliographystyle{abbrv}
\bibliography{dcc}

%\vfill

%\pagebreak

\appendix

\section*{Appendix}

\noindent 
We include full definitions of various systems described in this paper. (See next page.)

\vfill

\pagebreak

\begin{figure}[t]
\cenv{Typing rules (\dcc)}{}{
\vspace{-3mm}
\[ 
\left.
\begin{array}{lc}
{\sf (T\mbox{-}var)} & 
\infer
	{}
	{\Gamma, x : s, \Gamma'; \Pi \vdash x : s}
\\ \vspace{-3mm}
\\
{\sf (T\mbox{-}unit)} & 
\infer
	{}
	{\Gamma; \Pi \vdash \unit : \unitT}
\\ \vspace{-3mm}
\\
{\sf (T\mbox{-}abs)} & 
\infer
	{\Gamma, x : s; \Pi \vdash e : t}
	{\Gamma; \Pi \vdash \abs x e : (s \rightarrow t)}
\\  \vspace{-3mm}
\\
{\sf (T\mbox{-}app)} & \infer
	{\Gamma; \Pi \vdash e : s \rightarrow t \\ \Gamma; \Pi \vdash e' : s}
	{\Gamma; \Pi \vdash \app e {e'} : t}
\\ \vspace{-3mm}
\\
{\sf (T\mbox{-}pair)} & \infer
	{\Gamma; \Pi \vdash e_1 : s_1 \\ \Gamma; \Pi \vdash e_2 : s_2}
	{\Gamma; \Pi \vdash \tup {e_1} {e_2} : (s_1 \times s_2)}
\\  \vspace{-3mm}
\\
{\sf (T\mbox{-}proj)} & \infer
	{\Gamma; \Pi \vdash e : (s_1 \times s_2)}
	{\Gamma; \Pi \vdash \proj i e : s_i}
\\ \vspace{-3mm}
\\
{\sf (T\mbox{-}inj)} & \infer
	{\Gamma; \Pi \vdash e : s_i}
	{\Gamma; \Pi \vdash \inj i e : (s_1 + s_2)}
\\  \vspace{-3mm}
\\
{\sf (T\mbox{-}case)} & \infer
	{\Gamma; \Pi \vdash e : (s_1 + s_2) \\ \Gamma, x : s_i; \Pi \vdash e_i : s}
	{\Gamma; \Pi \vdash \split e x {e_1} {e_2} : s}
\\ \vspace{-3mm}
\\
{\sf (T\mbox{-}ret)} & \infer
	{\Gamma; \Pi \sqcup \ell \vdash e : s}
	{\Gamma; \Pi \vdash \strong \ell e : T_\ell (s)}
\\  \vspace{-3mm}
\\
{\sf (T\mbox{-}bind)} & \infer%*[sep=2mm]
	{\Gamma; \Pi \vdash e : T_\ell (s) \\ \Gamma, x : s; \Pi \vdash e' : t \\ \ell \preceq T_\Pi(t)}
	{\Gamma; \Pi \vdash \bind x {e} {e'} : t}
%\\ 
%{\sf (T\mbox{-}down)} & \infer*[sep=5mm]
%	{\Gamma; \Pi \vdash e : \Sigma_{\ell'} (s) \\ \ell \sqsubseteq \ell'}
%	{\Gamma; \Pi \vdash (\mathsf{downgrade}_\ell~e) : \Sigma_{\overline{\ell'}}(\Sigma_\ell(e))}
\end{array}
\right.
\]
}
%\vspace{-8mm}
\end{figure}

\begin{figure}[t]
\cenv{Indistinguishability relation (\dcc)}{}{
\vspace{-2mm}
\[
\left.
\hspace{-2mm}
\begin{array}{lc}
{\sf (I\mbox{-}unit)} &\!\! \infer
	{}
	{\unit \sim_\ell \unit : \unitT}
\\  \vspace{-3mm}
\\ 
{\sf (I\mbox{-}function)} &\!\! \infer
	{\forall e, e'.~~e \sim_\ell e' : s ~\Rightarrow~ \app v e \sim_\ell \app {v'} {e'} : t}
	{v \sim_\ell v' : (s \rightarrow t)}
\\  \vspace{-3mm}
\\ 
{\sf (I\mbox{-}product)} &\!\! \infer
	{e_1 \sim_\ell e'_1 : s_1 \\ e_2 \sim_\ell e'_2 : s_2}
	{\tup {e_1} {e_2} \sim_\ell \tup {e'_1} {e'_2} : (s_1 \times s_2)}
\\  \vspace{-3mm}
\\ 
{\sf (I\mbox{-}sum)} &\!\! \infer
	{e_i \sim_\ell e'_i : s_i}
	{\inj i {e_i} \sim_\ell \inj i {e'_i} : (s_1 + s_2)}
\\  \vspace{-3mm}
\\ 
{\sf (I\mbox{-}monad\mbox{-}1)} &\!\! \infer
	{\ell' \not\sqsubseteq \ell}
	{\strong {\ell'} e \sim_\ell \strong {\ell'} {e'} : T_{\ell'}(s)}
\\  \vspace{-3mm}
\\
{\sf (I\mbox{-}monad\mbox{-}2)} &\!\! \infer
	{e \sim_\ell e' : s}
	{\strong {\ell'} e \sim_\ell \strong {\ell'} {e'} : T_{\ell'}(s)}
\\  \vspace{-3mm}
\\ 
%{\sf (I\mbox{-}res)} & \infer
%	{e \sim_\ell e' : s}
%	{(\nu_{\ell'}~e) \sim_\ell (\nu_{\ell'}~e') : \mathcal B_{\ell'}(s)}
%\\ 
%{\sf (I\mbox{-}at)} & \infer
%	{v \sim_\ell v' : s}
%	{v \sim_\ell v' : s @ \ell}
%\\ 
{\sf (I\mbox{-}eval)} & \infer
	{e \longrightarrow^\star v \\ e' \longrightarrow^\star v' \\ v \sim_\ell v' : s}
	{e \sim_\ell e' : s}
\end{array}
\right.
\]
}
\vspace{-5mm}
\end{figure}
%\caption{}
%\label{fig:dccreln}
%\end{figure}

\begin{figure}[t]
\cenv{Typing rules (\dccd)}{}{
\vspace{-2mm}
\[
\left.
\begin{array}{lc}
{\sf (T^D\mbox{-}var)} & \infer
	{}
	{\Gamma, x : s, \Gamma'; \overline \Pi \vdash x : s}
\\ 
\vspace{-3mm} \\
{\sf (T^D\mbox{-}unit)} & \infer
	{}
	{\Gamma \vdash \unit : \unitT}
\\
\vspace{-3mm} \\
{\sf (T^D\mbox{-}abs)} & \infer
	{\Gamma, x : s; \overline \Pi \vdash e : t}
	{\Gamma; \overline \Pi \vdash \abs x e : (s \rightarrow t)}
\\ 
\vspace{-3mm} \\
{\sf (T^D\mbox{-}app)} & \infer
	{\Gamma; \overline \Pi \vdash e : s \rightarrow t \\ \Gamma; \overline \Pi \vdash e' : s}
	{\Gamma; \overline \Pi \vdash \app e {e'} : t}
\\
\vspace{-3mm} \\
{\sf (T^D\mbox{-}pair)} & \infer
	{\Gamma; \overline \Pi \vdash e_1 : s_1 \\ \Gamma; \overline \Pi \vdash e_2 : s_2}
	{\Gamma; \overline \Pi \vdash \tup {e_1} {e_2} : (s_1 \times s_2)}
\\ 
\vspace{-3mm} \\
{\sf (T^D\mbox{-}proj)} & \infer
	{\Gamma; \overline \Pi \vdash e : (s_1 \times s_2)}
	{\Gamma; \overline \Pi \vdash \proj  i e : s_i)}
\\
\vspace{-3mm} \\
{\sf (T^D\mbox{-}inj)} & \infer
	{\Gamma; \overline \Pi \vdash e : s_i}
	{\Gamma; \overline \Pi \vdash \inj i e : (s_1 + s_2)}
\\ 
\vspace{-3mm} \\
{\sf (T^D\mbox{-}case)} & \infer
	{\Gamma; \overline \Pi \vdash e : (s_1 + s_2)^\ell \\ \Gamma, x : s_i^\ell; \overline \Pi \vdash e_i : s}
	{\Gamma; \overline \Pi \vdash \split e x {e_1} {e_2} : s}
\\
%{\sf (T\mbox{-}ret)} & \infer
%	{\Gamma; \delta \sqcup \ell; \beta \sqcup \ell \vdash e : s}
%	{\Gamma; \overline \Pi \vdash (\eta_\ell~e) : \Sigma_\ell (s)}
%\\ 
%{\sf (T\mbox{-}bind)} & \infer*[sep=2mm]
%	{\Gamma; \overline \Pi \vdash e : \Sigma_\ell (s) \\ \Gamma, x : s; \overline \Pi \vdash e' : t \\ \delta \vdash \ell \preceq t}
%	{\Gamma; \overline \Pi \vdash (\mathsf{bind}~x = e~\mathsf{in}~e') : t}
%\\
\vspace{-3mm} \\
{\sf (T^D\mbox{-}ret)} & \infer
	{\Gamma; \overline \Pi \sqcup \ell \vdash e : s}
	{\Gamma; \overline \Pi \vdash \weak \ell e : \overline T_\ell (s)}
\\ 
\vspace{-3mm} \\
{\sf (T^D\mbox{-}bind)} & \infer
	{\Gamma; \overline \Pi \vdash e : \overline T_\ell (s) \\ \Gamma, x : s^\ell; \overline \Pi \vdash e' : t \\ \ell \leq \overline T_{\overline \Pi}(t)}
	{\Gamma; \overline \Pi \vdash \bind x e {e'} : t}
\end{array}
\right.
\]
}
\vspace{-5mm}
%\caption{}
%\label{fig:dccdtyping}
\end{figure}

\begin{figure*}
\cenvv{Leaking explicit flows via implicit flows: \dccd}{}{
\[
\left.
\begin{array}{rcl}
 \emph{leak}_\ell(\unitT) &  = & \abs {x: \overline T_\ell( \unitT )} \unit  \\
 \emph{leak}_\ell(s \times t) &  = & \abs {x: \overline T_\ell( s \times t )} 
ÊÊÊÊÊÊÊÊÊÊÊÊÊÊÊÊÊÊÊ\bind y x   \\
  && \quad {\tup {\emph{leak}_\ell(s) {\weak \ell {\proj 1 y}}} {\emph{leak}_\ell(t) {\weak \ell {\proj 2 y}}}} \\
 \emph{leak}_\ell((s + t)^{\ell'}) & = & \abs {x: \overline T_\ell( (s + t)^{\ell'} )}
ÊÊÊÊÊÊÊÊÊÊÊÊÊÊÊÊÊÊÊ\bind y x \\
  && \quad \mathsf{case}~y~\mathsf{of} \\
  && \qquad ~~~\mathsf{inj}_1(z_1).~{\inj 1 {(\emph{leak}_{\ell\sqcup \ell'}(s)Ê{\weak {\ell\sqcup \ell'} {z_1}} )}} \\
  && \qquad \|~\mathsf{inj}_2(z_2).~{\inj 2 {(\emph{leak}_{\ell\sqcup \ell'}(t) {\weak {\ell\sqcup \ell'} {z_2}} )}} \\
 \emph{leak}_\ell(s \rightarrow t) & = &  \abs {x: \overline T_\ell( s \rightarrow t )}
ÊÊÊÊÊÊÊÊÊÊÊÊÊÊÊÊÊÊÊ\\
  && \quad Ê\lambda z:s.~\bind f x ÊÊÊÊÊÊÊÊÊÊÊÊÊÊÊÊÊÊÊÊÊÊÊ{\emph{leak}_\ell(t) {\weak \ell {\app f z}}} \\
  \emph{leak}_\ell(\overline T_{\ell'} ( s )) & = & \abs {x: \overline T_\ell( \overline T_{\ell'} ( s ) )}
ÊÊÊÊÊÊÊÊÊÊÊÊÊÊÊÊÊÊÊÊÊÊÊÊÊ\bind y x \\
  && \quad {\weak {\ell'} {(\emph{leak}_{\ell'}(s) y)}}
\end{array}
\right.
\]
}
\end{figure*}

\begin{figure*}
\cenvv{Syntax: \dccdc}{}{
\emph{types} $s, t ::= \unitT \orelse (s \times t) \orelse (s + t) \orelse (s \rightarrow t) \orelse T_\ell(s) \orelse \overline T_\ell(s) \orelse s^\ell$ \\
\emph{terms} $e,v ::= \unit \orelse \tup e {e'} \orelse \proj i e \orelse \inj i e \orelse \split e x {e_1} {e_2} \\
~\quad\qquad\qquad \orelse \abs x e \orelse \app e {e'} \orelse \strong \ell e \orelse \weak \ell e \orelse \bind x e {e'} \orelse \weaken e$
}
%\end{figure*}
%\begin{figure}
\cenvv{Typing rules: \dccdc}{}{
\vspace{-2mm}
\[ 
\left.
\begin{array}{lc}
{\sf (T^{DC}\mbox{-}var)} & \infer
	{}
	{\Gamma, x : s, \Gamma'; \Pi; \overline \Pi \vdash x : s}
\\ 
{\sf (T^{DC}\mbox{-}unit)} & \infer
	{}
	{\Gamma; \Pi; \overline \Pi \vdash \unit : \unitT}
\\
{\sf (T^{DC}\mbox{-}abs)} & \infer
	{\Gamma, x : s; \Pi; \overline \Pi \vdash e : t}
	{\Gamma; \Pi; \overline \Pi \vdash \abs x e : (s \rightarrow t)}
\\ 
{\sf (T^{DC}\mbox{-}app)} & \infer
	{\Gamma; \Pi; \overline \Pi \vdash e : s \rightarrow t \\ \Gamma; \Pi; \overline \Pi \vdash e' : s}
	{\Gamma; \Pi; \overline \Pi \vdash \app e {e'} : t}
\\
{\sf (T^{DC}\mbox{-}pair)} & \infer
	{\Gamma; \Pi; \overline \Pi \vdash e_1 : s_1 \\ \Gamma; \Pi; \overline \Pi \vdash e_2 : s_2}
	{\Gamma; \Pi; \overline \Pi \vdash \tup {e_1} {e_2} : (s_1 \times s_2)}
\\ 
{\sf (T^{DC}\mbox{-}proj)} & \infer
	{\Gamma; \Pi; \overline \Pi \vdash e : (s_1 \times s_2)}
	{\Gamma; \Pi; \overline \Pi \vdash \proj i e : s_i}
\\
{\sf (T^{DC}\mbox{-}inj)} & \infer
	{\Gamma; \Pi; \overline \Pi \vdash e : s_i}
	{\Gamma; \Pi; \overline \Pi \vdash \inj i e : (s_1 + s_2)}
\\ 
{\sf (T^{DC}\mbox{-}case)} & \infer
	{\Gamma; \Pi; \overline \Pi \vdash e : (s_1 + s_2)^\ell \\  \Gamma, x : s_i^\ell; \Pi; \overline \Pi \vdash e_i : s}
	{\Gamma; \Pi; \overline \Pi \vdash \split e x {e_1} {e_2} : s}
\\
{\sf (T^{DC}\mbox{-}ret\mbox{-}1)} & \infer
	{\Gamma; \Pi \sqcup \ell; \overline \Pi \sqcup \ell \vdash e : s}
	{\Gamma; \Pi; \overline \Pi \vdash \strong \ell e : T_\ell (s)}
\\ 
{\sf (T^{DC}\mbox{-}ret\mbox{-}2)} & \infer
	{\Gamma; \Pi; \overline \Pi  \sqcup \ell \vdash e : s}
	{\Gamma; \Pi; \overline \Pi \vdash \weak \ell e : \overline T_\ell (s)}
\\ 
{\sf (T^{DC}\mbox{-}bind\mbox{-}1)} & \infer*[sep=5mm]
	{\Gamma; \Pi; \overline \Pi \vdash e : T_\ell (s) \\ \Gamma, x : s; \Pi; \overline \Pi \vdash e' : t \\ \ell \preceq T_\Pi(t)}
	{\Gamma; \Pi; \overline \Pi \vdash \bind x {e} {e'} : t}
\\ 
{\sf (T^{DC}\mbox{-}bind\mbox{-}2)} & \infer*[sep=5mm]
	{\Gamma; \Pi; \overline \Pi \vdash e : \overline T_\ell (s) \\ \Gamma, x : s^\ell; \Pi; \overline \Pi \vdash e' : t \\ \ell \leq T_\Pi(t)}
	{\Gamma; \Pi; \overline \Pi \vdash \bind x {e} {e'} : t}
\\ 
{\sf (T^{DC}\mbox{-}weaken)} & \infer
	{\Gamma; \Pi; \overline \Pi \vdash e : T_\ell(s)}
	{\Gamma; \Pi; \overline \Pi \vdash \weaken e : T_{\beta(\ell)}(\overline T_\ell (s))}
%\\ 
%{\sf (T^{DC}\mbox{-}down)} & \infer*[sep=5mm]
%	{\Gamma; \Pi \vdash e : \overline \Pi_{\ell'} (s) \\ \ell \sqsubseteq \ell'}
%	{\Gamma; \Pi \vdash (\mathsf{downgrade}_\ell~e) : \overline \Pi_{\overline{\ell'}}(\overline \Pi_\ell(e))}
\end{array}
\right.
\]
}
%\vspace{-15mm}
%\caption{}
%\label{fig:dccdctyping}
%\end{figure*}
\cenvv{Protection rules: \dccdc}{}{
\vspace{-3mm}
\[
\left.
\begin{array}{rl}
{\sf (P\mbox{-}unit)}~ & {\ell \preceq \unitT}
\\ 
{\sf (P\mbox{-}product)}~ & \inferx
	{\ell \preceq s \andalso \ell \preceq t}
	{\ell \preceq (s \times t)}
\\ 
{\sf (P\mbox{-}function)}~ & \inferx
	{\ell \preceq t}
	{\ell \preceq (s \rightarrow t)}
\\
%{\sf (P\mbox{-}mon)}~ & \inferx
%	{\ell' \vdash \ell \preceq s}
%	{\ell \preceq T_{\ell'} (s)}
%\\ 
{\sf (P\mbox{-}monad\mbox{-}1,2)}~ & \inferx
	{\ell \sqsubseteq \ell'}
	{\ell \preceq T_{\ell'}(s)}
\quad,\quad \inferx
	{\ell \preceq s}
	{\ell \preceq T_{\ell'}(s)}
\\
{\sf (P\mbox{-}effect)}~ & \inferx
	{\ell \preceq s}
	{\ell \preceq \overline T_{\ell'}(s)}
\\
{\sf (P^D\mbox{-}unit)}~ & {\ell \leq \unitT}
\\ 
{\sf (P^D\mbox{-}product)}~ & \inferx
	{\ell \leq s \andalso \ell \leq t}
	{\ell \leq (s \times t)}
\\ 
{\sf (P^D\mbox{-}function)}~ & \inferx
	{\ell \leq t}
	{\ell \leq (s \rightarrow t)}
\\
%{\sf (S\mbox{-}enc)}~ & \inferx
%	{\ell' \vdash \ell \leq s}
%	{\ell \leq \mathcal B_{\ell'}(s)}
%\\ 
%{\sf (S\mbox{-}mon)}~ & \inferx
%	{\ell \leq \mathcal B_{\ell'}(s)}
%	{\ell \leq \Sigma_{\ell'}(s)}	
%\\ 
{\sf (P^D\mbox{-}effect)}~ & \inferx
	{\ell \sqsubseteq \ell'}
	{\ell \leq \overline T_{\ell'}(s)}	
\quad,\quad \inferx 
	{\ell \leq s}
	{\ell \leq \overline T_{\ell'}(s)}	
\\
{\sf (P^D\mbox{-}sum)}~ & \inferx
	{\ell \leq s \andalso \ell \leq t}
	{\ell \leq (s + t)}
\\
{\sf (P^D\mbox{-}monad)}~ & \inferx
	{\ell \leq \overline T_{\ell'}(s)}
	{\ell \leq T_{\ell'}(s)}
\end{array}
\right.
\]
}
\end{figure*}

\begin{figure*}
\cenvv{Open type equations: \dccdc}{}{
\vspace{-2mm}
\[
\left.
\begin{array}{rl}
%{\sf (E\mbox{-}res)} & \infer
%	{\ell \sqsubseteq \ell'}
%	{\vdash s@\ell < s @ \ell'}
%\\ 
{\sf (E\mbox{-}open\mbox{-}1,2)}~ & 
	{(s^\ell)^{\ell'} = s^{\ell \sqcup \ell'}}
\quad,\qquad 
	{s = s^\bot}	
\\
{\sf (E\mbox{-}unit)}~ & 
	{\unitT^\ell = \unitT}
\\ 
{\sf (E\mbox{-}product)}~ & 
	{(s \times t)^\ell = (s^\ell \times t^\ell)}
\\ 
{\sf (E\mbox{-}function)}~ & 
	{(s \rightarrow t)^\ell = s \rightarrow t^\ell}
\\ 
{\sf (E\mbox{-}effect\mbox{-}1,2)}~ & 
	{\overline T_{\ell'}(s)^\ell = \overline T_{\ell'}(s^\ell)}
\quad,\quad \inferx
	{\ell \sqsubseteq \ell'}
	{\overline T_{\ell'}(s)^\ell = \overline T_{\ell'}(s)}
\\
{\sf (E\mbox{-}monad)}~ & \inferx
	{\overline T_{\ell'}(s)^\ell =  \overline T_{\ell'}(s^\ell)}
	{ T_{\ell'}(s)^\ell =  T_{\ell'}(s^\ell)}
%\\ 
%{\sf (E\mbox{-}mon)}~ & 
%	{\Sigma_{\ell'}(s)@\ell = \Sigma_{\ell'}(s @ \ell)}
%\quad,\qquad \inferx
%	{\ell \sqsubseteq \ell'}
%	{\Sigma_{\ell'}(s)@\ell = \Sigma_{\ell'}(s)}
\end{array}
\right.
\]
}
\cenvv{Syntax: \dcccd}{}{
\emph{types} $s, t ::= \unitT \orelse (s \times t) \orelse (s + t) \orelse (s \rightarrow t) \orelse T_\ell(s) \orelse \overline T_\ell(s) \orelse s^\ell$ \\
\emph{terms} $e,v ::= \unit \orelse \tup e {e'} \orelse \proj i e \orelse \inj i e \orelse \split e x {e_1} {e_2}$\\
$~~\quad\qquad\qquad \orelse \abs x e \orelse \app e {e'} \orelse \strong \ell e \orelse \bind x e {e'}$
}
%\begin{figure}
\cenvv{Typing rules: \dcccd}{}{
\vspace{-2mm}
\[ 
\left.
\begin{array}{lc}
{\sf (T^{CD}\mbox{-}var)} & \infer
	{}
	{\Gamma, x : s, \Gamma'; \Pi; \Sigma \vdash x : s}
\\ 
{\sf (T^{CD}\mbox{-}unit)} & \infer
	{}
	{\Gamma; \Pi; \Sigma \vdash \unit : \unitT}
\\
{\sf (T^{CD}\mbox{-}abs)} & \infer
	{\Gamma, x : s; \Pi; \Sigma \vdash e : t}
	{\Gamma; \Pi; \Sigma \vdash \abs x e : (s \rightarrow t)}
\\ 
{\sf (T^{CD}\mbox{-}app)} & \infer
	{\Gamma; \Pi; \Sigma \vdash e : s \rightarrow t \\ \Gamma; \Pi; \Sigma \vdash e' : s}
	{\Gamma; \Pi; \Sigma \vdash \app e {e'} : t}
\\
{\sf (T^{CD}\mbox{-}pair)} & \infer
	{\Gamma; \Pi; \Sigma \vdash e_1 : s_1 \\ \Gamma; \Pi; \Sigma \vdash e_2 : s_2}
	{\Gamma; \Pi; \Sigma \vdash \tup {e_1} {e_2} : (s_1 \times s_2)}
\\ 
{\sf (T^{CD}\mbox{-}proj)} & \infer
	{\Gamma; \Pi; \Sigma \vdash e : (s_1 \times s_2)}
	{\Gamma; \Pi; \Sigma \vdash \proj i e : s_i}
\\
{\sf (T^{CD}\mbox{-}inj)} & \infer
	{\Gamma; \Pi; \Sigma \vdash e : s_i}
	{\Gamma; \Pi; \Sigma \vdash \inj i e : (s_1 + s_2)}
\\ 
{\sf (T^{CD}\mbox{-}case)} & \infer*[sep=5mm]
	{\Gamma; \Pi; \Sigma \vdash e : (s_1 + s_2)^\ell \\  \ell \not\sqsubseteq \bot \Rightarrow \Sigma \not\sqsubseteq \ell \\ \Gamma, x : s_i^\ell; \Pi; \Sigma \vdash e_i : s}
	{\Gamma; \Pi; \Sigma \vdash \split e x {e_1} {e_2} : s}
\\
{\sf (T^{CD}\mbox{-}ret)} & \infer
	{\Gamma; \Pi \sqcup \ell; \Sigma \vdash e : s}
	{\Gamma; \Pi; \Sigma \vdash \strong \ell e : T_\ell (s)}
\\ 
{\sf (T^{CD}\mbox{-}bind\mbox{-}1)} & \infer*[sep=5mm]
	{\Gamma; \Pi; \Sigma \vdash e : T_\ell (s) \\ \Gamma, x : s; \Pi; \Sigma \vdash e' : t \\ \ell \preceq T_\Pi(t)}
	{\Gamma; \Pi; \Sigma \vdash \bind x {e} {e'} : t}
\\ 
{\sf (T^{CD}\mbox{-}bind\mbox{-}2)} & \infer*[sep=5mm]
	{\Gamma; \Pi; \Sigma \vdash e : T_\ell (s) \\ \Gamma, x : s^\ell; \Pi; \Sigma \sqcap \ell \vdash e' : t \\ \ell \leq T_\Pi(t)}
	{\Gamma; \Pi; \Sigma \vdash \bind x {e} {e'} : t}
%\\ 
%{\sf (T^{CD}\mbox{-}down)} & \infer*[sep=5mm]
%	{\Gamma; \Pi \vdash e : \Sigma_{\ell'} (s) \\ \ell \sqsubseteq \ell'}
%	{\Gamma; \Pi \vdash (\mathsf{downgrade}_\ell~e) : \Sigma_{\overline{\ell'}}(\Sigma_\ell(e))}
\end{array}
\right.
\]
}
\cenvv{Protection rules and Open type equations: \dcccd}{}{
\mbox{Same as those for \dccdc.}
}
\end{figure*}
%

%\vspace{-15mm}
%\caption{}
%\label{fig:dcccdtyping}
%\end{figure}

\end{document}